\documentclass{article}

\usepackage{PRIMEarxiv}

\usepackage[utf8]{inputenc} 
\usepackage[T1]{fontenc}    
\usepackage{hyperref}       
\usepackage{url}            
\usepackage{booktabs}       
\usepackage{amsfonts}       
\usepackage{nicefrac}       
\usepackage{microtype}      
\usepackage{lipsum}
\usepackage{fancyhdr}       
\usepackage{graphicx}       
\graphicspath{{media/}}     

\title{
\thanks{\textit{\underline{Citation}}: 
\textbf{Authors. Title. Pages.... DOI:000000/11111.}} 
}

\title{Positive and negative extinction of active particles}
\author{
  Ari Sihvola, Henrik Wall\'en, Pasi Yl\"a-Oijala \\
  Department of Electronics and Nanoengineering \\
  Aalto University \\
  Finland\\
  \texttt{\{ari.sihvola,henrik.wallen,pasi.yla-oijala\}@aalto.fi} \\
   \And
  Rashda Parveen \\
  Department of Electronics \\
  Quaid-i-Azam University \\
  Pakistan\\
  \texttt{rashda@ele.qau.edu.pk} \\
  }

\usepackage{amsmath}
\usepackage{graphicx}
\usepackage{subfig}
\usepackage{epsfig}

\newcommand\jj{\text{j}}
\newcommand\E{\varepsilon}

\begin{document}

\maketitle

\begin{abstract}
Using analytical Lorenz--Mie scattering formalism and numerical methods, we analyze the response of active particles to electromagnetic waves.  The particles are composed of homogeneous, non-magnetic, and dielectrically isotropic medium. Spherical scatterers and sharp and rounded cubes are treated. The absorption cross-section of active particles is negative, thus showing gain in their electromagnetic response. Since the scattering cross-section is always positive, their extinction can be either positive, negative, or zero. We construct a five-class categorization of active and passive dielectric particles. We point out the enhanced backscattering phenomenon that active scatterers display, and also discuss extinction paradox and optical theorem. Finally, using COMSOL Multiphysics and an in-house Method-of-Moments code, the effects of the non-sphericity of active scatterers on their electromagnetic response are illustrated. 
\end{abstract}

\section{Introduction}

The effect of electric and magnetic fields on material media leads to charges and currents which, in the linear response domain, are proportional to the excitation. For the mathematical analysis of the interaction between waves and matter, these effects are usually condensed into so-called constitutive relations between the field vectors and flux densities. These relations are not always straightforward. During the present century, along with the development of metamaterials, the richness and complexity of possible electromagnetic and optical responses for materials has expanded enormously to include anisotropic, chiral, non-reciprocal, gyrotropic, magnetoelectric, and several other effects \cite{Capolino}.

When particles and objects composed of homogeneous materials are exposed to electromagnetic waves, also their shape, size, and other geometric parameters play a role in their global response. Even very simple homogeneous scatterers may display optically fascinating and complicated responses, as is witnessed by the full complexity of a rainbow that is a result of interaction of natural light with simple spherical water droplets. 

In the present study, we focus on the electromagnetic response of particles that are characterized by active material properties. In other words, unlike water or other naturally existing media which are dissipative and passive, the medium displays gain. The problem to be treated consists of a scatterers which are non-magnetic, isotropic, and homogeneous, characterized by a uniform scalar (complex) permittivity. The objects to be analyzed are symmetric: spheres and cubes with varying rounding on their edges and corners. For the analysis of spherical scatterers, we apply the Mie theory, while for the non-spherical shapes, numerical approaches are used.

Mie scattering (also known as Lorenz--Mie scattering \cite{Lorenz,Mie}) is indeed a very classical theory to analyze, in a full-wave extension, the scattering, absorption, and extinction due to spherical particle exposed to electromagnetic and optical radiation. While there are, in the literature, several extensions of Mie theory to, for example, core--shell particles \cite{kerker-layered-1951}, layered structures \cite{Hightower:88}, and spheres whose surface is defined by boundary conditions \cite{Henrik-MI-2011,RSL2019}, it seems that spheres with active response have not been extensively documented. Some early theoretical studies on the electromagnetic response of active scatterers exist from 1970s \cite{Alexopoulos_active,Kerker_active,Kerker_active79}, and also worth seeing is the later study on Kerker conditions \cite{PhysRevLett.125.073205}. The mechanisms to generate active responses (via external energy pumping and stimulated emission lasing) can be enhanced by strong resonance effects in nanoparticles with silver shell and gain-impregnated silica core \cite{Gordon-Ziolkowski}, as well as quantum-dot active coated nanoparticles \cite{6062374}, where in the analysis Maxwell Garnett homogenization turns out to be useful \cite{Holmstrom,Campbell-Ziolkowski}.

In the following, we define an active dielectric medium by the sign of the imaginary part of its permittivity. Following the time-harmonic notation $\exp(\mathrm{j}\omega t)$, the complex relative permittivity reads $\E'-\jj \E''$ which means that a positive $\E''$ corresponds to lossy (dissipative) media, negative $\E''$ to gainy (active) media, and $\E''=0$ stands for lossless (and gainless) media \cite[p.~91]{methods}. Gain materials are active while passive media $\E''\geq0$ comprise both the dissipative and lossless classes. The characterization of active and passive scatterers leads to an interesting classification of particles where we distinguish them based on the signs of their extinction cross section and the imaginary part of their permittivity. We show how the behavior of the multipolar electric and magnetic Mie coefficients is strongly dependent on the degree of activity, emphasize the strong backscattering enhancement for gainy particles, and revisit the fundamental concepts of extinction paradox and optical theorems. Mie series results are verified with numerical approaches. Finally, these numerical approaches are used to study how the shape of active scattering particles affects their electromagnetic and optical response.

\section{Lorenz--Mie scattering by active spheres}

The classical Lorenz--Mie analysis of the scattering process when an isotropic dielectric sphere in free space is exposed to an electromagnetic wave leads to the scattering (sca), absorption (abs), and extinction (ext) cross sections of the scattering object. Here we follow the notation in the textbook by Bohren and Huffman~\cite{Bohren-Huffman}. Normalized by the geometric cross section of the sphere, the cross sections become dimensionless efficiencies which can be computed from the following series
\begin{align}
    \label{eq:Qsca}
    Q_\text{sca} & =\frac{2}{x^2} \sum_{n=1}^{\infty }(2n+1)
                    \left(\left| a_n \right|^2 + \left| b_n \right|^2\right)\\
    \label{eq:Qext}
    Q_\text{ext} & =\frac{2}{x^2}{\sum_{n=1}^{\infty }}(2n+1)\operatorname{Re}\left\{a_n + b_n\right\}\\
   \label{eq:Qabs}
    Q_\text{abs} & =Q_\text{ext}-Q_\text{sca}
\end{align}
The electric and magnetic Mie coefficients $a_n$ and $b_n$ appearing in these expressions read, as functions of the relative permittivity $\E$ and radius $a$ of the sphere, as
  \begin{align}
    a_n &= \frac{\sqrt{\varepsilon}\,\psi_n(\sqrt{\E}\,x)\psi_n'(x) 
      - \psi_n(x)\psi_n'(\sqrt{\E}\,x)}
    {\sqrt{\E}\,\psi_n(\sqrt{\E}\,x)\xi_n'(x)
      - \xi_n(x)\psi_n'(\sqrt{\E}\,x)}  \\
    b_n &= \frac{\psi_n(\sqrt{\E}\,x)\psi_n'(x)
      - \sqrt{\E}\,\psi_n(x)\psi_n'(\sqrt{\E}\,x)}
    {\psi_n(\sqrt{\E}\,x)\xi_n'(x)
      - \sqrt{\E}\,\xi_n(x)\psi_n'(\sqrt{\E}\,x)} 
   \end{align}
where the dimensionless size parameter is $x=2\pi a/\lambda$, with $\lambda$ being the wavelength. Here, the Riccati--Bessel functions $\psi_n, \xi_n$ are defined in terms of the ordinary spherical Bessel $(j_n)$ and Hankel $(h_n)$ functions:
\begin{equation}
  \psi_n(\rho) = \rho\,j_n(\rho), \qquad
  \xi_n(\rho) = \rho\,h_n^{(2)}(\rho), 
\end{equation}

The validity of the Mie expansions has been proven for dielectric, magnetic, lossless, and lossy spheres, as well as impedance-boundary scatterers for a myriad of applications during the past decades. However, due to the fact that not so many studies have been published for Mie scattering of active spheres, care has to be taken when the parametric space of the material character of the sphere is extended by changing the sign of the imaginary part of the permittivity. This is the focus of the present section where we look carefully on the validity of the Mie scattering analysis for active scattering spheres. In addition to comparing the analytical Mie results against numerical schemes, we check the results against the known convergence behavior of the efficiency series for non-active scatterers.

\subsection{Wiscombe criterion and convergence of the Mie series}

Since the scattering, absorption, and extinction efficiencies \eqref{eq:Qsca}--\eqref{eq:Qabs} are formally infinite series, they need to be truncated in the numerical evaluation. Electrically (optically) large spheres require more terms than small ones: the larger the size parameter $x$, the more terms are needed to maintain the accuracy. The widely-used rule in the existing literature is the so-called Wiscombe criterion \cite{Wiscombe}: the required minimum number of terms in the Mie series is
        \begin{equation}\label{eq:W}
       N = \left\{ \begin{array}{ll}
       \left\lceil x + 4x^{1/3}+1\right\rceil & x\leq 8 \\ & \\
       \left\lceil x + 4.05x^{1/3}+2\right\rceil & 8<x<4200 \\ & \\
       \left\lceil x + 4x^{1/3}+2\right\rceil & 4200\leq x<20{,}000 
       \end{array}\right.
        \end{equation}
where the symbol of ceiling $\left\lceil \cdots \right\rceil$ denotes the closest integer larger than the argument.

\begin{figure}[htbp]
\centering
\includegraphics[width=0.7\textwidth]{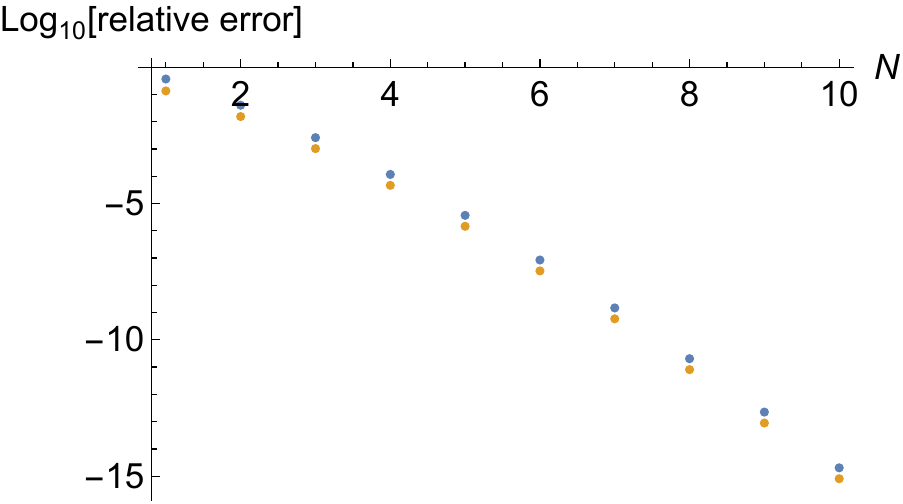} 
\caption{\label{fig:W1} The relative error (the vertical axis shows the number of correct digits of the approximation) of the extinction efficiency $Q_\mathrm{ext}$, as function of the number of terms $N$ in the Mie expansion for $x=2$. Blue dots: passive scatterer ($\E=2-\mathrm{j}$); orange dots: active scatterer ($\E=2+\mathrm{j}$).}
\end{figure}

To illustrate the convergence of the Mie series and the validity of the Wiscombe criterion, Figure~\ref{fig:W1} shows the error of a truncated series for the extinction efficiency as function of the number of terms, for passive ($\E=2-\jj$) and active ($\E=2+\jj$) spheres with size parameter $x=2$. The figure shows that the Wiscombe criterion is extremely conservative when computing the extinction efficiency: for $x=2$, the Wiscombe criterion \eqref{eq:W} requires 9 terms in the series, which, according to Figure~\ref{fig:W1}, gives already 13 correct digits! Another interesting observation is that with a given number of terms $N$, the computed efficiency of the active sphere is more accurate than for the dissipative one, at least in this case.
        
\begin{figure}[htbp]
\centering
\subfloat[$x=2.14$]{\includegraphics[width=0.45\textwidth]{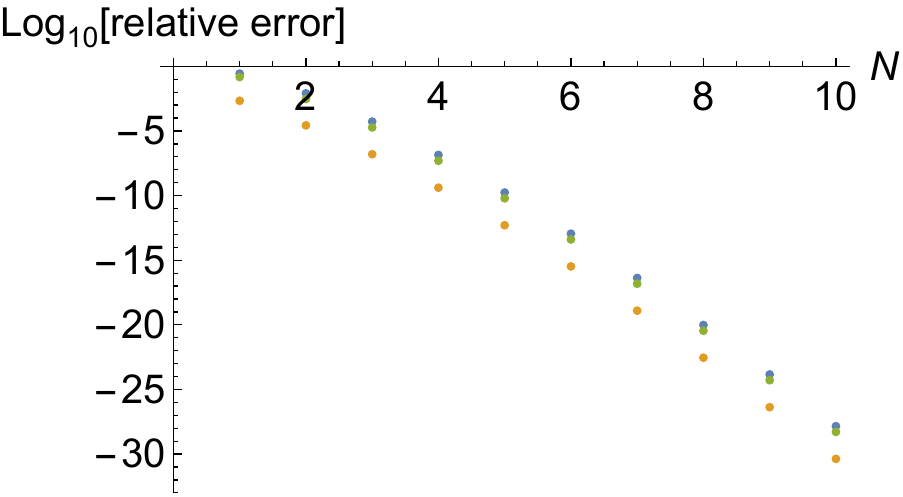}}
\hspace{3mm}
\subfloat[$x=0.2,\E=-2.1$]{\includegraphics[width=0.45\textwidth]{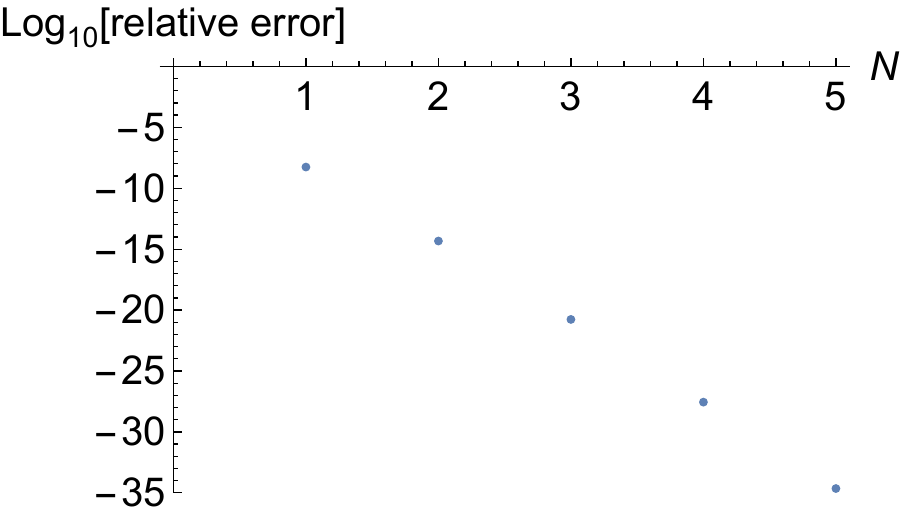}}
\caption{\label{fig:W2} 
The relative error (the vertical axis shows the number of correct digits of the approximation) of the scattering efficiency $Q_\mathrm{sca}$, as function of the number of terms $N$ in the Mie expansion. (a) Blue dots: passive scatterer ($\E=2-\mathrm{j}$); orange dots: active scatterer ($\E=2+\mathrm{j}$), green dots: lossless scatterer ($\E=2$). (b) The same rapid convergence for a lossless plasmonic resonance (negative-permittivity sphere).}
\end{figure}

One would expect that the convergence of the Mie series would be slower when the efficiency is very large, like at resonances, rather than at this arbitrary chosen case ($x=2,\E=2\mp\jj$). Figure~\ref{fig:W2} shows the corresponding comparison at the first resonance of the active sphere ($x=2.14,\E=2+\jj$) and the conjugate dissipative sphere ($x=2.14,\E=2-\jj$). For this case, the efficiency of the active case converges much more rapidly than in Figure~\ref{fig:W1}: for a given number of terms, the active sphere is three orders of magnitude more accurate than the dissipative (or lossless, $\E = 2$) case. For comparison, also shown is the fast convergence of the series for a lossless plasmonic (negative-permittivity) sphere.

\begin{figure}[htbp]
\centering
\includegraphics[width=0.7\textwidth]{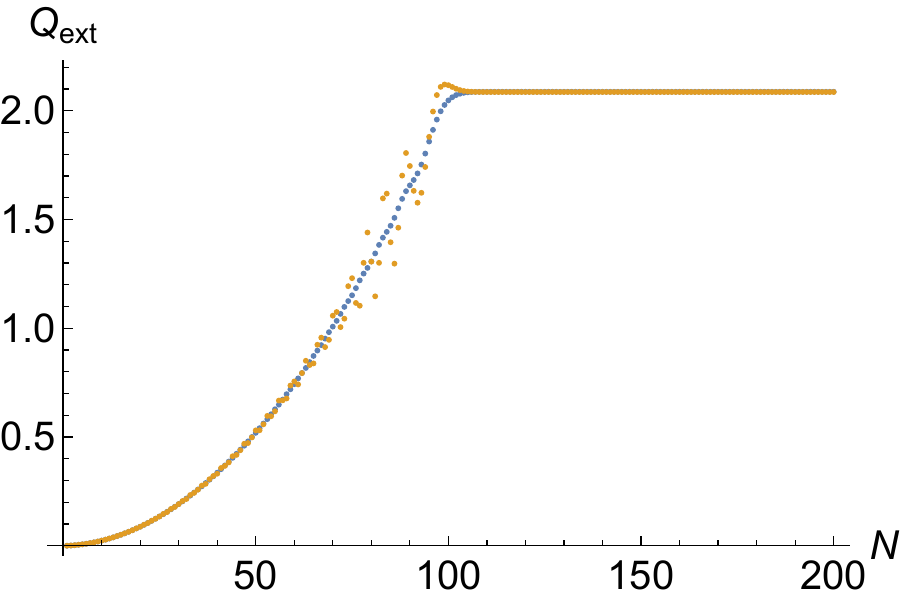} 
\caption{\label{fig:W5} The convergence of the Mie series as the number of terms $N$ for the extinction efficiency of passive (blue dots, $\E=2-\jj$) and active (orange dots, $\E=2+\jj$) large spheres (the size parameter $x=100$).}
\end{figure}

Another look at the convergence is given in Figure~\ref{fig:W5} where the extinction efficiency estimate of large spheres ($x=100$) is shown as the number of terms $N$ in the Mie series increases. For this size, the Wiscombe criterion requires 121 terms, while in the figure, the curves stabilize already at around 100 terms. Extinction is very close but not equal for the passive ($\E=2-\jj$) and active ($\E=2+\jj$) cases:  $Q_\text{ext}=2.086$ and $Q_\text{ext}=2.087$, respectively. Active and passive spheres seem to converge with similar speeds, but the convergence behavior of the passive one (blue dots) is more stable as $N$ increases.
  
\subsection{Extinction paradox}
    
\begin{figure}[htbp]
\centering
\subfloat[$\E=2\mp\mathrm{j}$]{\includegraphics[width=0.45\textwidth]{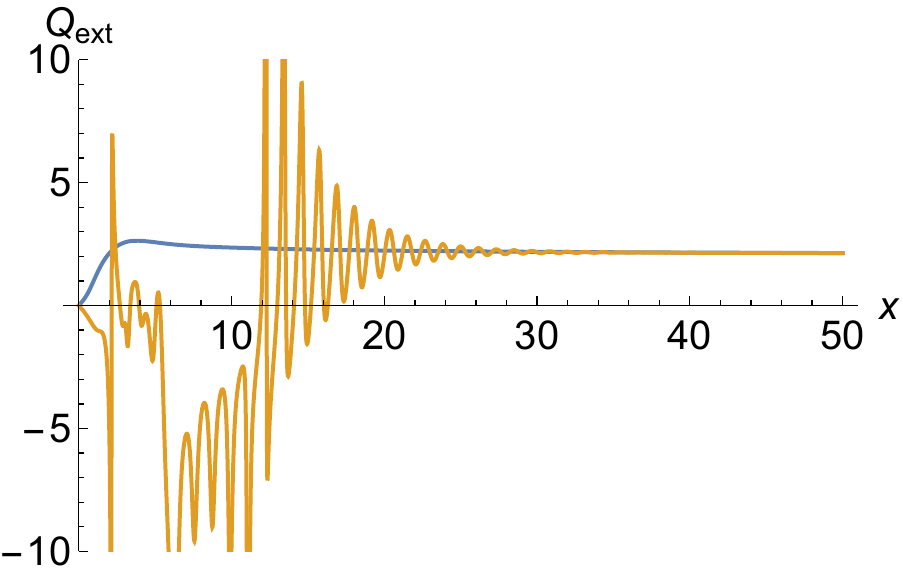}}
\hspace{3mm}
\subfloat[$\E=2\mp\mathrm{j}10$]{\includegraphics[width=0.45\textwidth]{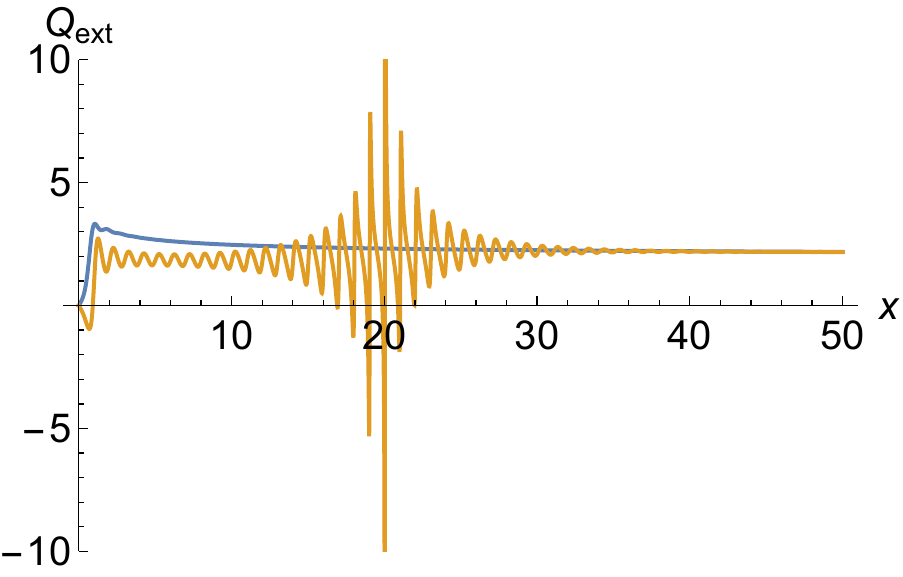}}
\caption{\label{fig:ext-par} The extinction efficiency of active and passive spheres, for a wide variation of size parameter $x$. Blue curves are for passive and orange ones for active scatterers.}
\end{figure}

A strong theoretical result for the scattering problem is the so-called extinction paradox \cite{Brillouin,Kristensson}, according to which in the high-frequency limit, the extinction cross section of the particle equals twice its geometrical cross section. In other words, the extinction efficiency approaches the value $2$ when $x$ becomes large. This result has been validated for passive and perfectly conducting spheres, for which $Q_\text{ext}$ converges to this value rather smoothly. Figure~\ref{fig:ext-par} corroborates the extinction paradox also in the active side: the extinction efficiency of two cases of active spheres ($\E=2+\jj$ and $\E=2+\jj 10$) are shown as function of $x$. The figure shows that despite the fact that the extinction efficiency displays strong oscillations for moderate size parameters (compared with the dissipative case), the curves still stabilize close to the value $2$ when $x$ reaches values of around $40$.
    
\subsection{Verification with (FEM/MoM) computations}

In this section, the Mie solutions for an active sphere are verified with numerical solutions. We use an in-house Method-of-Moments (MoM) code (based on \cite{Pasi03,Pasi05}) and commercial software COMSOL Multiphysics~\cite{comsol} which is based on the finite element method (FEM). Both methods are available for arbitrarily shaped particles.

In MoM, an electromagnetic scattering problem is reformulated as a surface integral equation (SIE) for equivalent electric $\mathbf{J}$ and magnetic $\mathbf{M}$ surface current densities. This yields surface mesh and surface unknowns, and the field behavior in a medium is described with a Green's function. Compared to the methods based on volume meshing, such as FEM, these properties could be particularly beneficial in active media where the fields inside the medium may have strong variations. 

\begin{figure}[htbp]
\centering
  \subfloat[]{\includegraphics[width=0.43\textwidth]{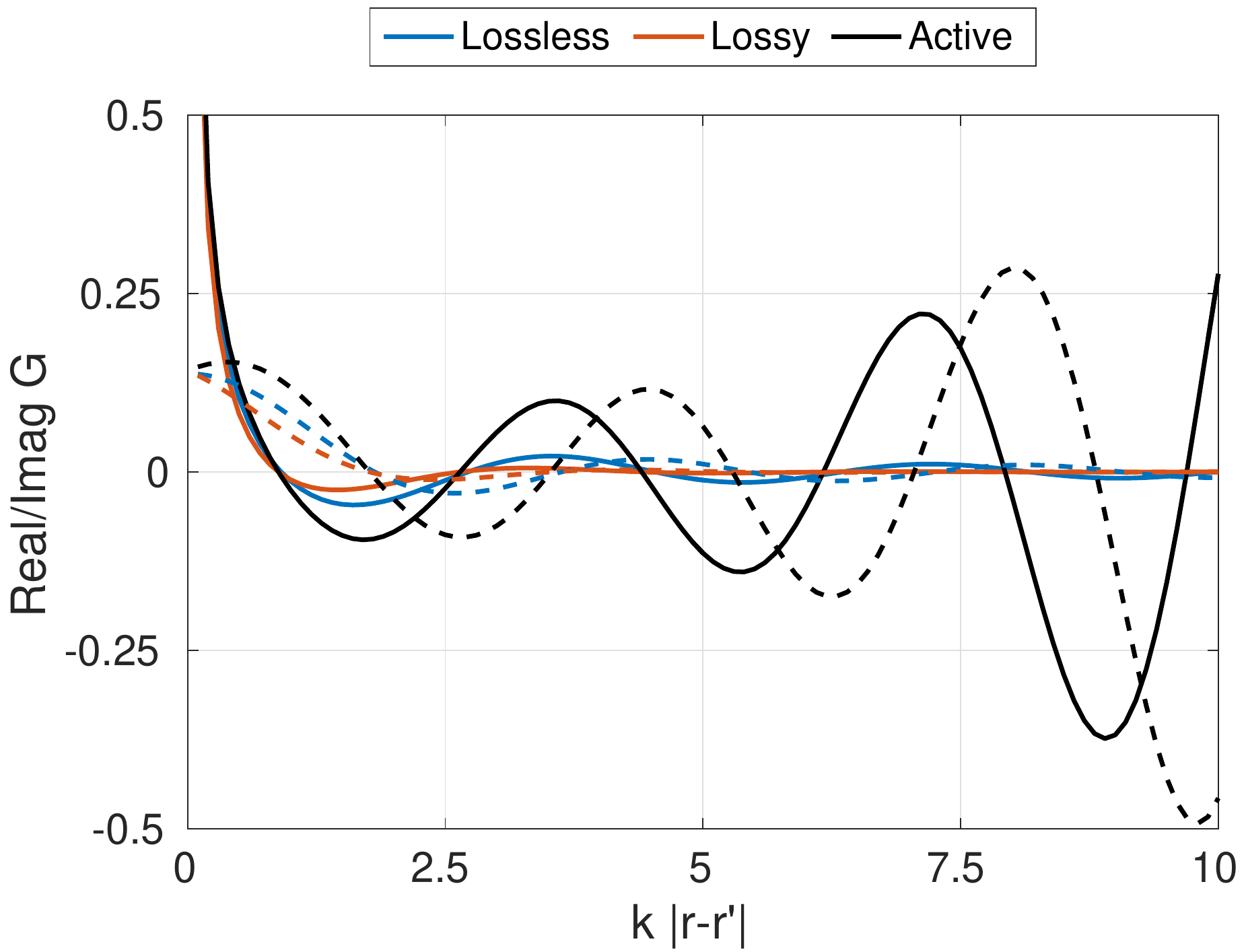}}\quad
  \subfloat[]{\includegraphics[width=0.43\textwidth]{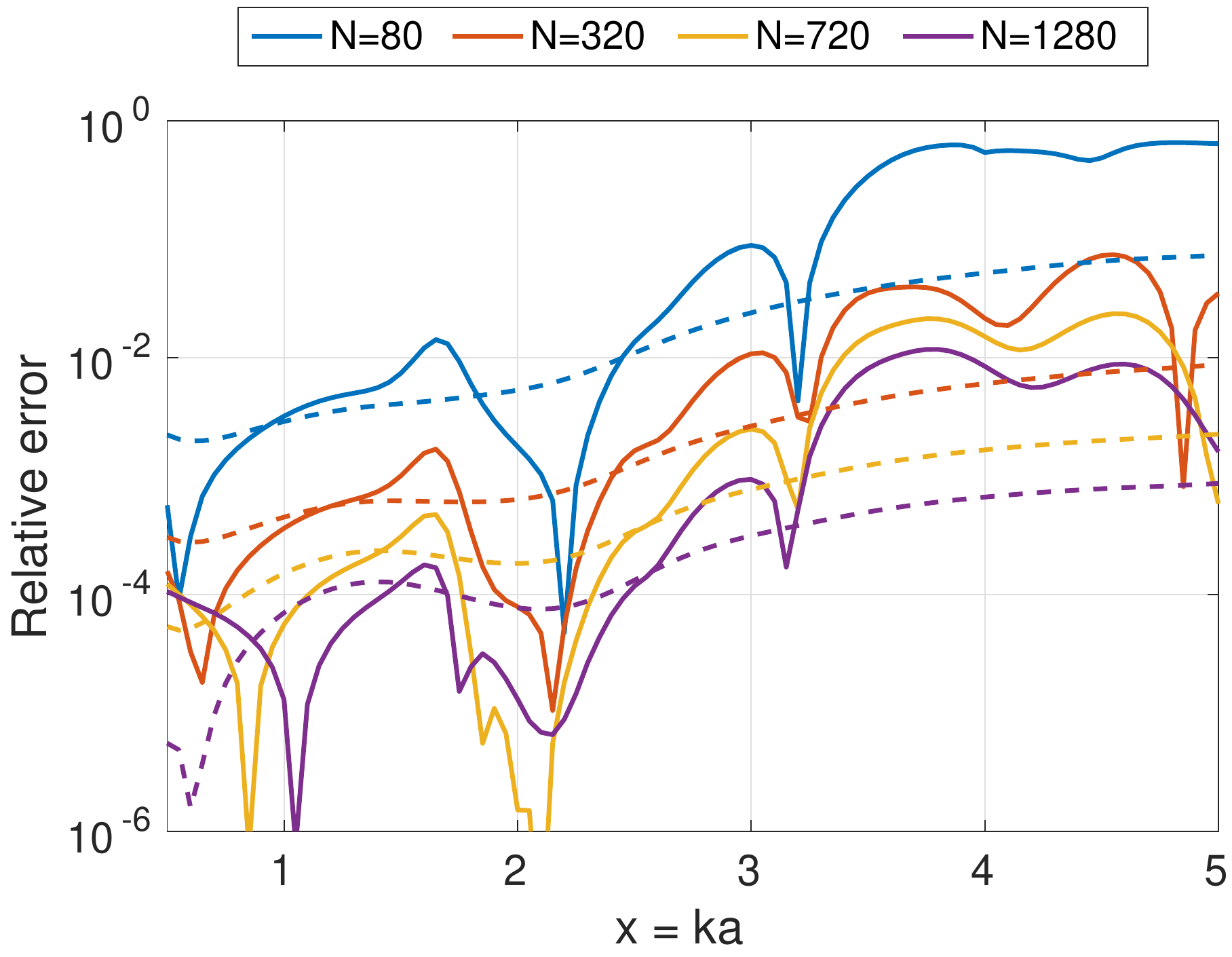}}
\caption{\label{fig:Greenfunction} (a) Real (solid lines) and imaginary (dashed lines) parts of the Green's function in lossless, ($\E = 3$), dissipative ($\E = 3 - 1.5\jj$) and active ($\E = 3 + 1.5\jj)$ media. (b) Relative error of the MoM solution for a passive sphere (dashed lines, $\E = 3 - 1.5\jj$) and active sphere (solid lines, $\E = 3 + 1.5\jj)$ with increased number of planar triangular surface elements $(N)$.}
\end{figure}

In this work the classical Poggio--Miller--Chang--Harrington--Wu--Tsai (PMCHWT) \cite{Poggio73} formulation with Rao--Wilton--Glisson (RWG) \cite{Rao82} basis and test functions is applied. The Green's function is defined by
\begin{equation}
G(\mathbf{r},\mathbf{r}') = \frac{e^{-\jj k R}}{4\pi R} = \frac{1}{4\pi R} \left( \cos{kR} - \jj \sin{kR}\right),\;\;\; R = |\mathbf{r} - \mathbf{r}'|
\end{equation}
with an active medium wave number $(\E'' <0)$
\begin{equation}
\label{eq:wavenumber}
    k = \omega \sqrt{\E\E_0 \mu_0} = \omega \sqrt{\E' - \jj \E^{''}} \sqrt{\E_0\mu_0}.
\end{equation}
 The behavior of the Green's function in lossless ($\E = 3$), dissipative ($\E = 3 - 1.5\jj$) and active ($\E = 3 + 1.5\jj)$ media is illustrated in Figure~\ref{fig:Greenfunction}(a). In all three cases, the real part of the Green's function has $1/R$ type singularity as the field $\mathbf{r}$ and source $\mathbf{r}'$ points coincide, while the imaginary parts are regular as $R\rightarrow 0$. This indicates that similar singularity subtraction \cite{Wilton84} or cancellation \cite{Khayat05} techniques that are developed for the numerical evaluation of the Green's function in passive and lossless media would work also in the active case. The difference is that in active media the Green's function, both the real and imaginary parts, is a strongly oscillating function which amplitude increases as the distance between the field and source points increases. In passive dissipative medium the Green's function is decaying function as $R$ increases. 
 
Convergence of the MoM solution versus the number of planar triangular elements is studied in Figure \ref{fig:Greenfunction}(b) for a passive (lossy) and an active sphere. In both cases the same analytical and numerical methods \cite{Pasi03} with the same number of integration points are used to evaluate the singular integrals involving the Green's function and its gradient. To obtain the same accuracy as in the passive case, higher mesh density is required in the active case, particularly as the size parameter $x$ increases.

\begin{figure}[htbp]
    \centering
    \includegraphics[width=.4\textwidth]{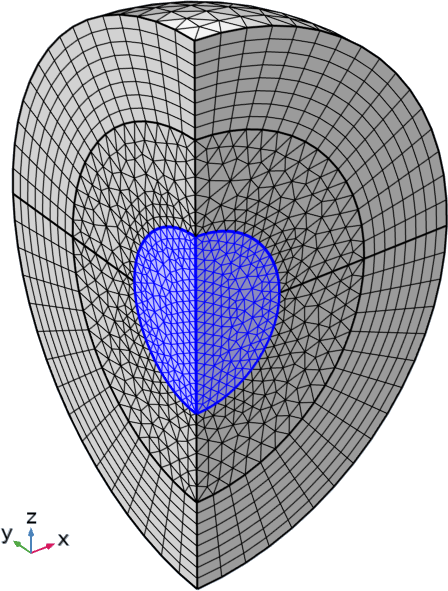}
    \caption{Computational geometry in COMSOL for a sphere with relative permittivity $\E=3\pm1.5\jj$ and size parameter $x=3$. The sphere (blue) and the surrounding air is meshed using a free tetrahedral mesh while the PML region has a swept mesh. The incident plane wave propagates in the $+z$ direction and is $x$-polarized so that we have the perfect-electric-conductor symmetry in the $yz$ plane and the perfect-magnetic-conductor symmetry in the $xz$ plane.}
    \label{fig:comsol_sphere_3_mesh}
\end{figure}

In COMSOL~\cite{comsol}, the computational setup is fairly similar to the verification example \texttt{rcs\textunderscore{}sphere} that is included with the RF Module. The computational geometry is shown in Figure~\ref{fig:comsol_sphere_3_mesh} for one particular case. To ensure good accuracy, we have maximum edge length $\lambda/10$ in the free tetrahedral meshes and an 8-layer swept mesh in the spherical perfectly matched layer (PML). Both the air layer and the PML are $\lambda_0/2$ thick. The scattering efficiency $Q_\text{sca}$ is computed using a surface integral of the time average power outflow of the scattered field, while the absorption efficiency $Q_\text{abs}$ is computed using a volume integral of the electromagnetic power loss density in the sphere. 

\begin{figure}[htbp]
\centering
  \subfloat[$Q_\text{sca}, Q_\text{abs}$]{\includegraphics[width=0.43\textwidth]{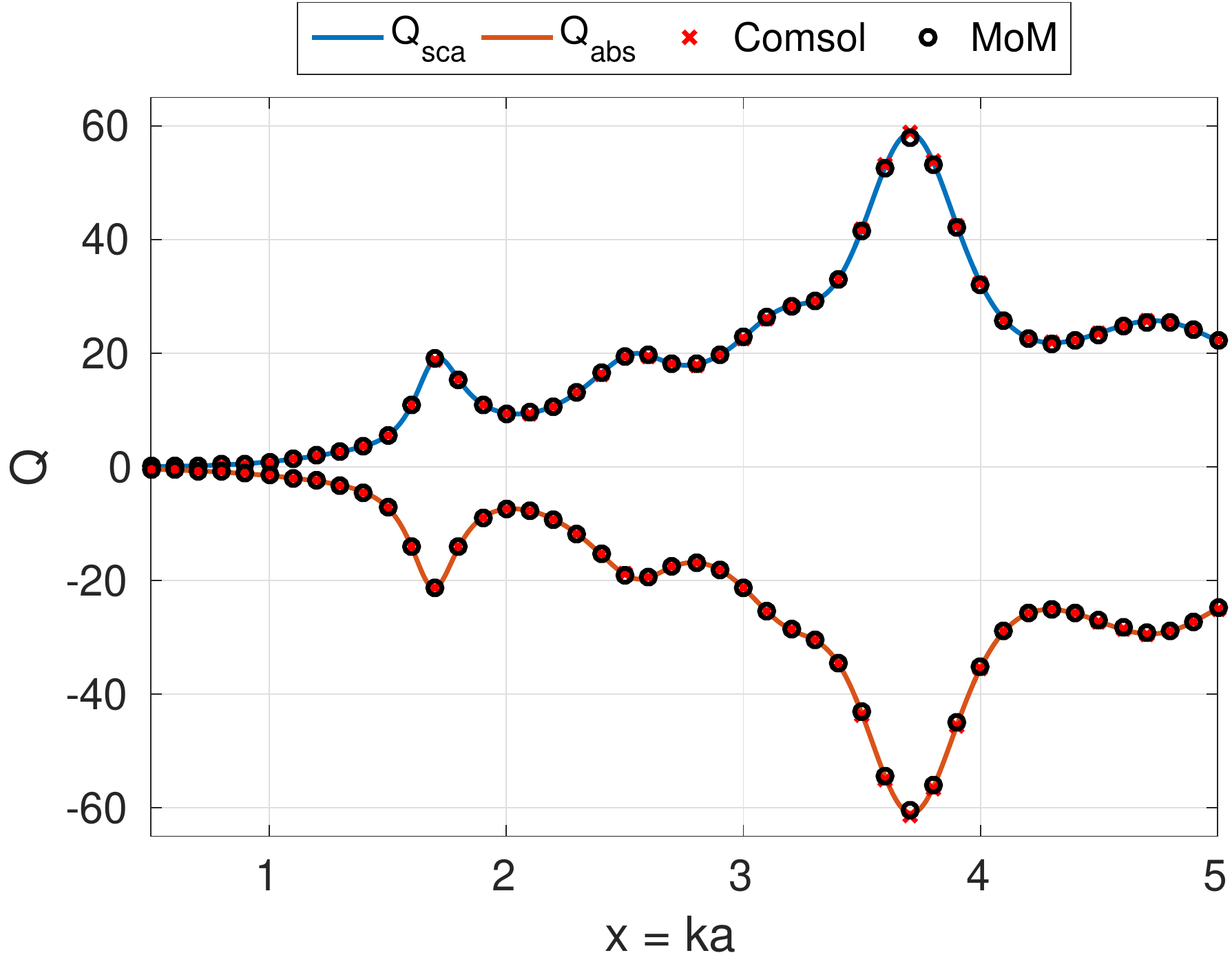}}\quad
  \subfloat[rel. error]{\includegraphics[width=0.43\textwidth]{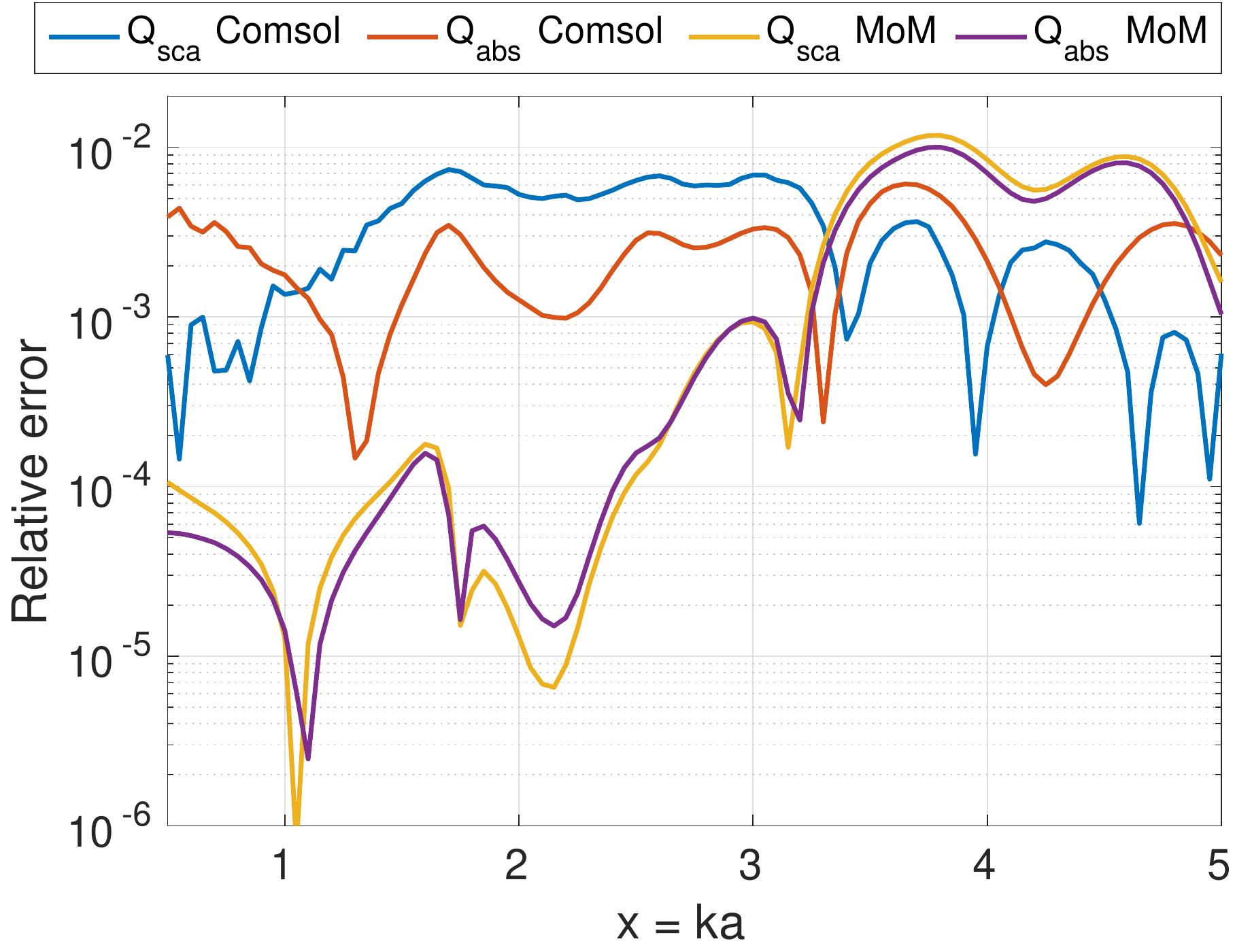}}
\caption{\label{fig:comparison} (a) Scattering and absorption efficiency with Mie series, COMSOL and in-house MoM code versus the size parameter $x=ka$. (b) Relative errors of COMSOL and MoM solutions for $Q_\text{sca}$ and $Q_\text{abs}$. An active sphere with $\E = 3+1.5\jj$.}
\end{figure}

\begin{figure}[htbp]
\centering
  \subfloat[$Q_\text{ext}$]{\includegraphics[width=0.43\textwidth]{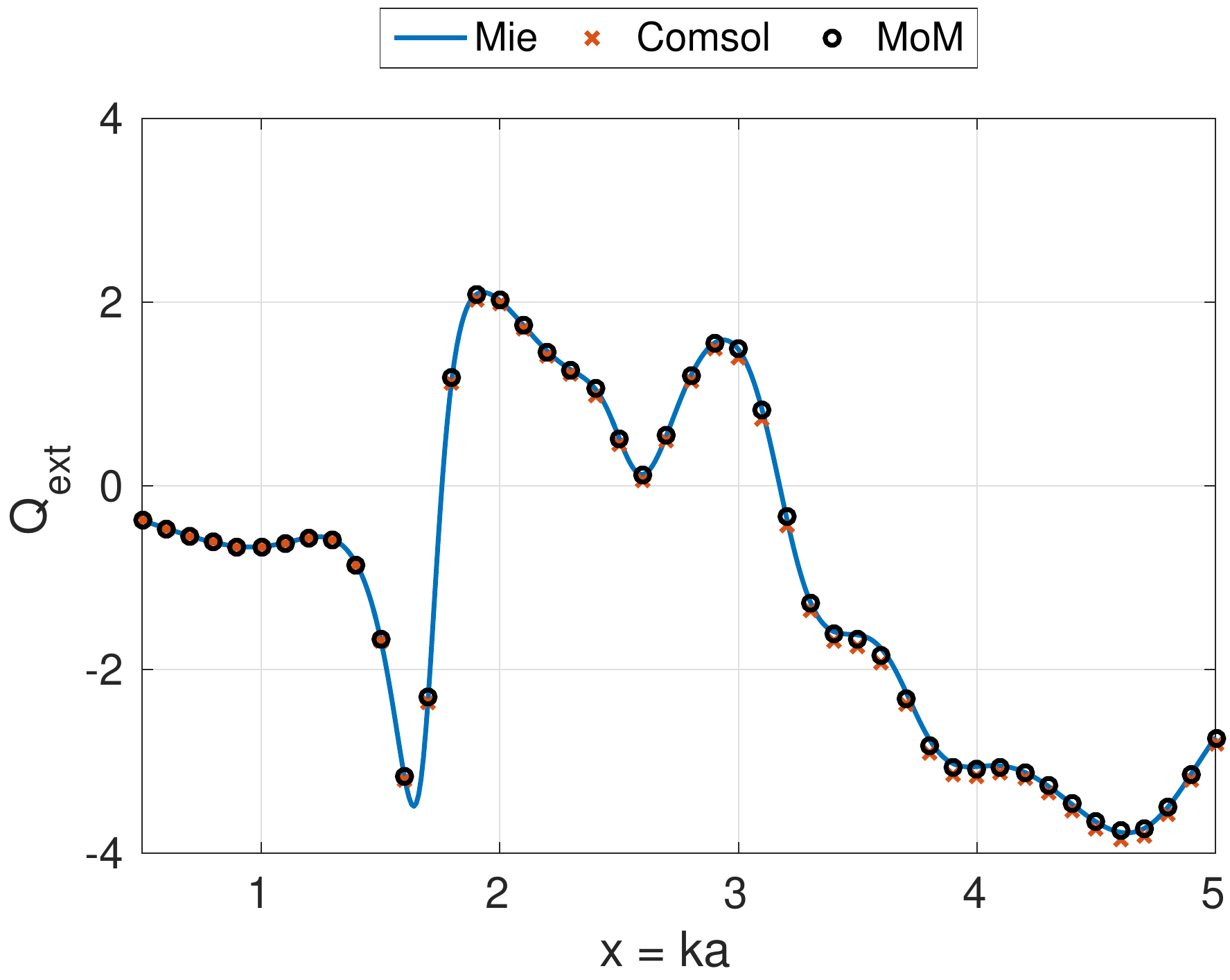}}\quad
  \subfloat[abs. error]{\includegraphics[width=0.43\textwidth]{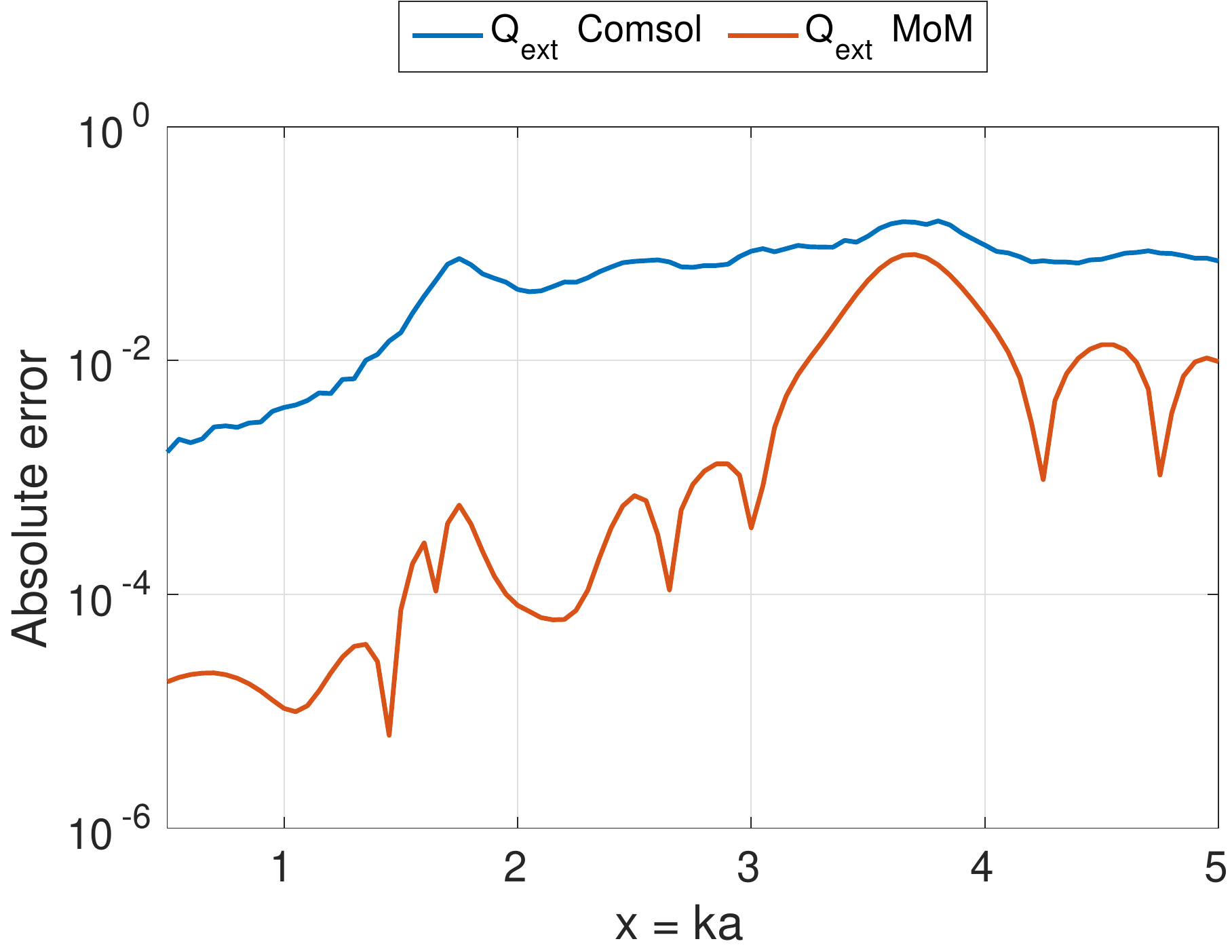}}
\caption{\label{fig:comparison2} (a) Extinction efficiency with Mie series, COMSOL and in-house MoM code. (b) Absolute errors of the COMSOL and MoM solutions for $Q_\text{ext}$. An active sphere with $\E = 3+1.5\jj$.}
\end{figure}

A comparison of the efficiencies of an active sphere computed with Mie series, MoM code, and COMSOL software is shown in Figures \ref{fig:comparison} and \ref{fig:comparison2}. In the MoM solution the mesh is the same for all size parameters, while in COMSOL the thickness of the air layer and the PML relative to the sphere radius depend on the size parameter $x$ and so also the mesh depends on $x$. Good accuracy is obtained with both numerical methods and they agree well with the Mie series results.

\section{Mie coefficients and backscattering enhancement}

The behavior of the electric and magnetic Mie coefficients $a_n$ and $b_n$ determine the spectral characteristics of the scattering and extinction of the spheres with a given permittivity. It turns out that there is a drastic difference between the coefficients and consequently the global response of active dielectric spheres compared to passive and dissipative ones.

\subsection{Properties of Mie coefficients}

\begin{figure}[htbp]
\centering
\subfloat[$a_1$]{\includegraphics[width=0.45\textwidth]{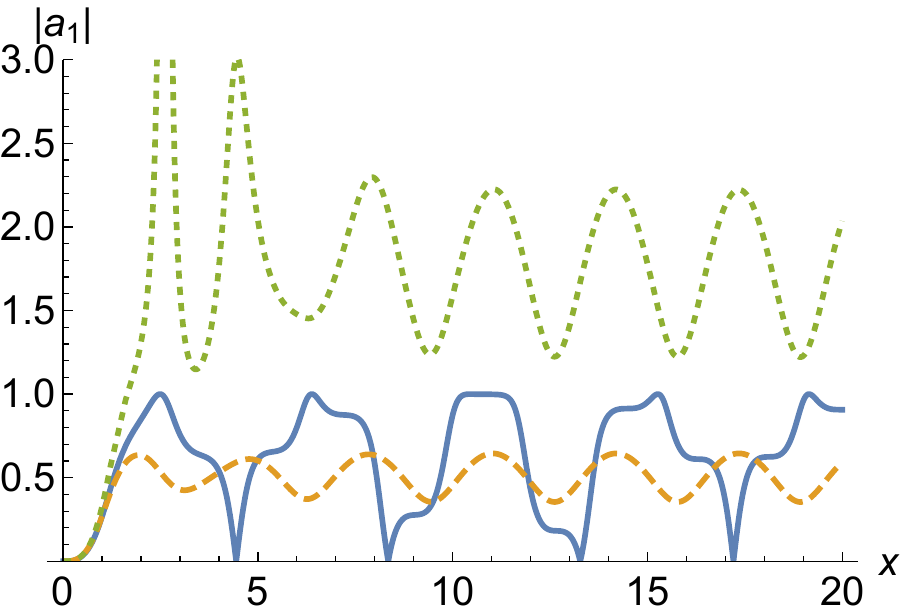}} 
\hspace{3mm}
\subfloat[$b_1$]{\includegraphics[width=0.45\textwidth]{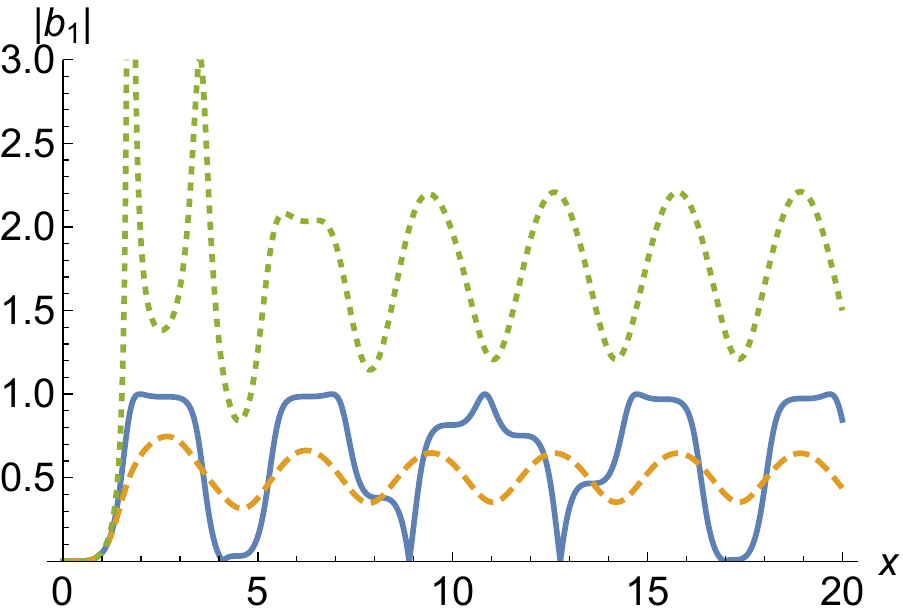}}
\caption{\label{fig:ab1} Comparison of Mie coefficients as function of the size parameter $x$ for a lossless ($\E=3$, solid blue), dissipative ($\E=3-\jj $, dashed orange), and active  ($\E=3+ \jj $, dotted green) non-magnetic sphere.}
\end{figure}

For lossless/gainless scatterers ($\E''=0$), the Mie coefficients satisfy 
\begin{equation}\label{eq:lossless}
\text{Re}\{c_n\} = \left|c_n\right|^2
\end{equation}
where $c_n$ stands for both the electric ($a_n$) and magnetic ($b_n$) coefficients. From \eqref{eq:lossless}, it can be seen that the absolute value of the coefficients is always between $0$ and $+1$. For dissipative scatterers, the maximum value is less than one, while for active scatterers, the coefficients do not have an upper limit. A comparison of Mie coefficients (Figure~\ref{fig:ab1}) for the three types of scatterers illustrates these properties.

An interesting result for the Mie coefficients (valid for all $x,n$, and $\E=\E'-\jj\E''$) is the following:
\begin{equation}\label{eq:uusitulos}
    2c_n(\E)c_n^*(\E^*) = c_n(\E)+c_n^*(\E^*)
\end{equation}
where the conjugate of a complex number is denoted by $\E^*=\left(\E'-\jj\E''\right)^*=\E'+\jj\E''$.
For lossless scatterers $(\E''=0)$, Equation~\eqref{eq:uusitulos} returns the result \eqref{eq:lossless}. This formula \eqref{eq:uusitulos} provides a straightforward way to compute the Mie coefficients of an active sphere from the corresponding passive ones.

\subsection{Multipolar contributions to the efficiencies}

\begin{figure}[htbp]
\centering
\subfloat[]{\includegraphics[width=0.45\textwidth]{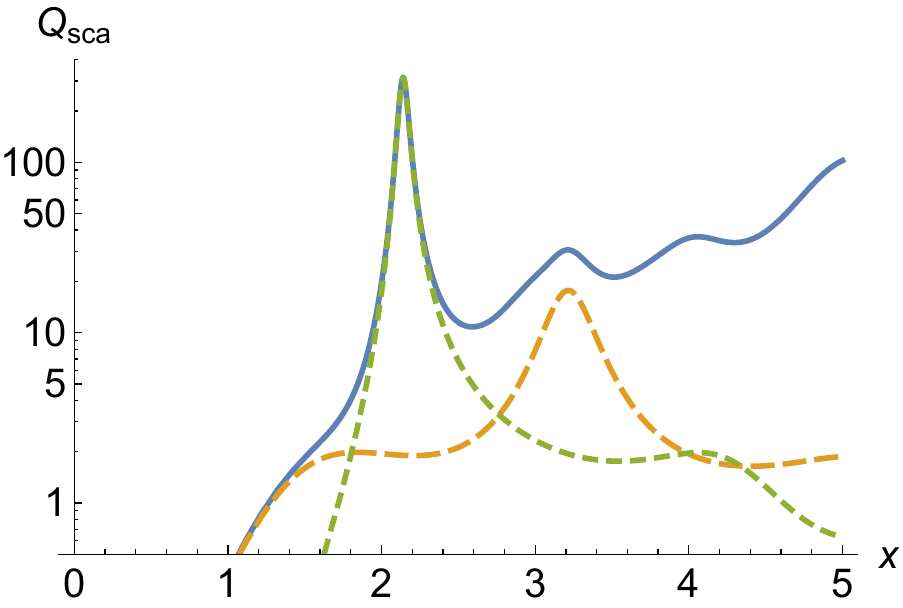}}
\hspace{3mm}
\subfloat[]{\includegraphics[width=0.45\textwidth]{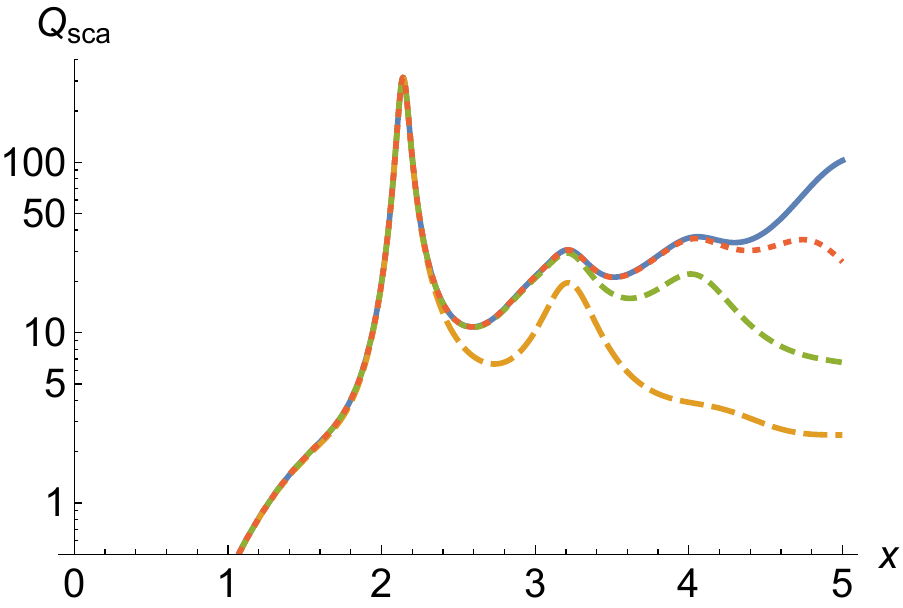}}
\caption{\label{fig:partial} (a) The contributions of the electric dipole scattering corresponding to the coefficient $a_1$ (long-dashed orange) and the magnetic dipole from coefficient $b_1$ (short-dashed green) into the total scattering efficiency $Q_\mathrm{sca}$ (solid blue), as function of the size parameter $x$. (b) The contributions of the orders of multipoles: both dipoles ($n=1$; long-dashed orange), dipoles and quadrupoles ($n=1,2$; short-dashed green), and dipoles, quadrupoles, and hexapoles ($n=1,2,3$; dotted red). The results are for an active sphere with $2+\jj$.}
\end{figure}

As an example of how the different multipoles contribute to the scattering of a sphere, consider an active sphere with relative permittivity $\E=2+\jj$ as function of its size. Figure~\ref{fig:partial} displays how the various electric and magnetic multipoles account for the total scattering efficiency $Q_\text{sca}$. It is noteworthy that the main, rather sharp, resonance at around $x=2.1$ is due to the magnetic dipole while the other resonances from higher-order modes remain softer and take place for optically larger spheres. This is indeed dramatically different from the behavior of passive scatterers for which the magnetic dipole resonance cannot be distinguished in the scattering cross section for scatterers with relative permittivity as low as $\E'=2$.
    
\begin{figure}[htbp]
\centering
\subfloat[]{\includegraphics[width=0.45\textwidth]{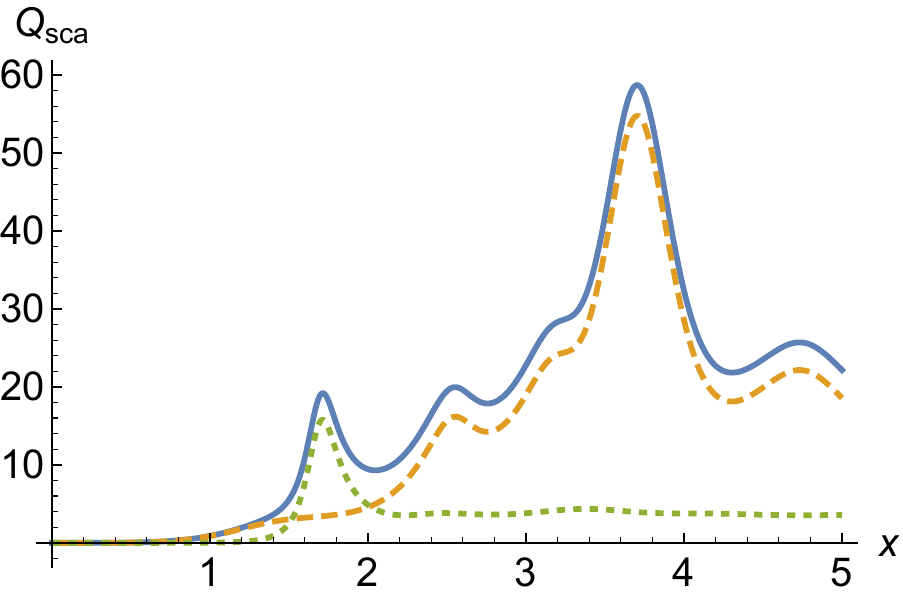}}
\hspace{3mm}
\subfloat[]{\includegraphics[width=0.45\textwidth]{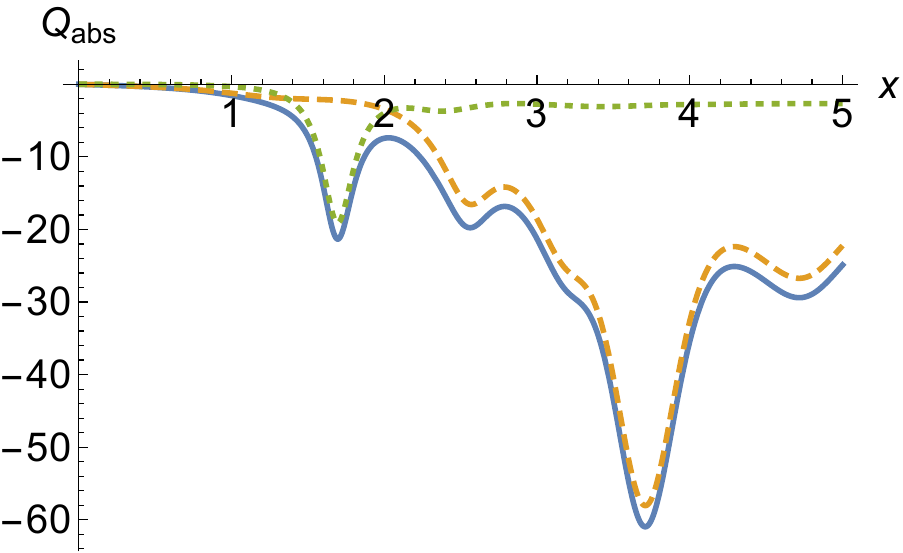}}
\caption{\label{fig:dipoles} (a) The contribution of electric multipoles (orange dashed) and magnetic multipoles (green dotted) to the scattering efficiency (blue solid) as function of the size parameter $x$. (b) The same for the absorption efficiency. The results are for an active sphere with $\E= 3+\jj 1.5$.}
\end{figure}

Another perspective how the electric and magnetic multipoles contribute to scattering is shown in Figure~\ref{fig:dipoles}. There the contribution of electric and magnetic multipoles into the scattering and absorption efficiencies of an active sphere with $\E=3+\jj 1.5$ is illustrated separately. Note the fact that, concerning the magnetic multipoles, only the magnetic dipole $b_1$ has a notable effect; all other resonances in scattering and absorption efficiencies arise from the electric multipoles $a_n$.
    
 
 \subsection{Backscattering enhancement} 

\begin{figure}[htbp]
\centering
\subfloat[backward-to-forward ratio]{\includegraphics[width=0.49\textwidth]{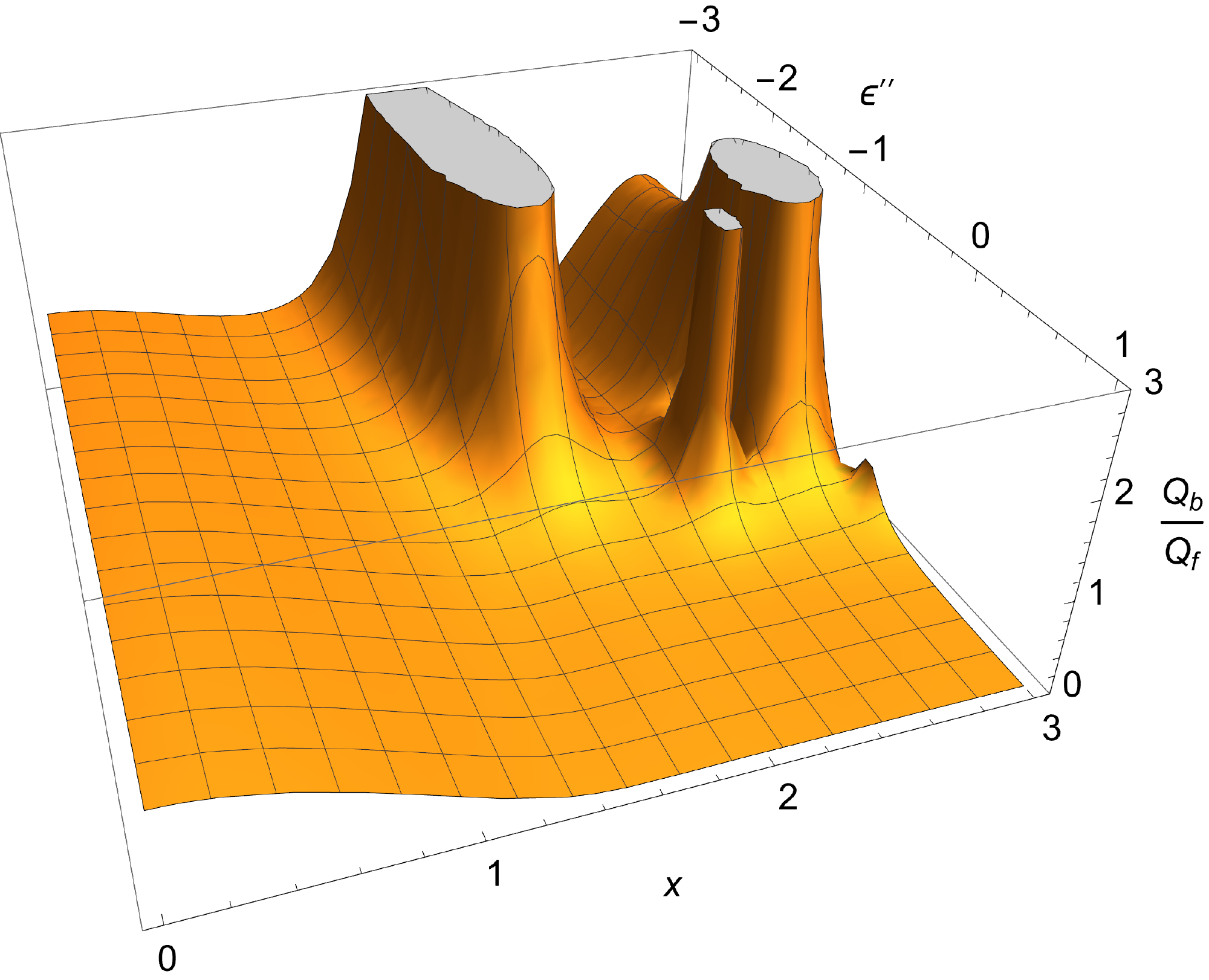}}
\hspace{1mm}
\subfloat[asymmetry parameter]{\includegraphics[width=0.49\textwidth]{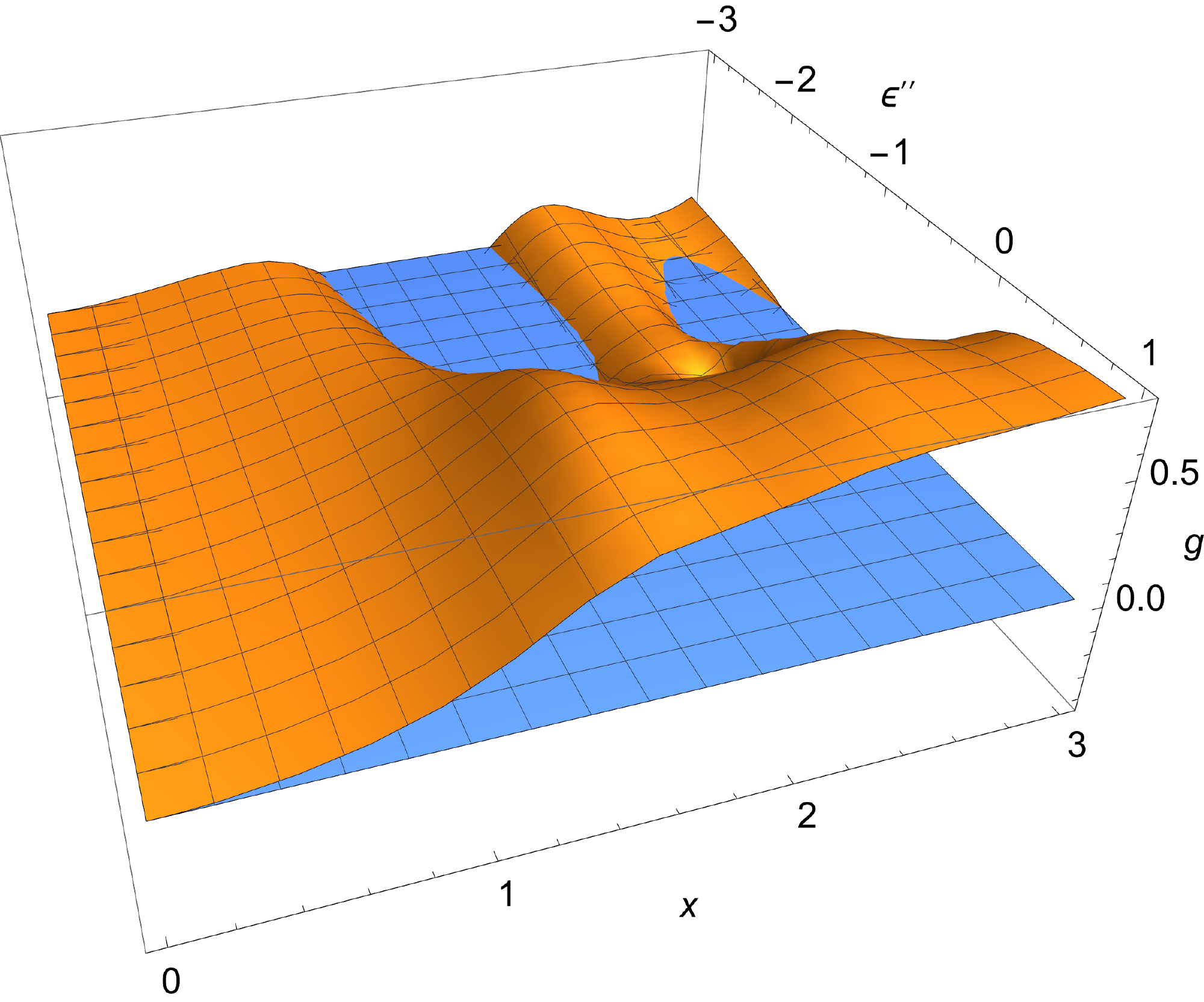}}
\caption{\label{fig:g}Illustration of the balance between backscattering and forward scattering of active $\E''<0$ and dissipative $\E''>0$ spheres, with real part of the relative permittivity being $\E'=3$, as function of the size parameter $x$ and the imaginary part $\E''$. The blue plane shows the level $g=0$.}
\end{figure}

 In our previous study, we have emphasized the particular character in the scattering behavior of active dielectric objects: the enhanced backscattering \cite{activeRSL2021}. The (monostatic) radar cross section of the sphere may reach very high values. From the Mie coefficients, the backscattering and forward scattering cross sections and efficiencies ($Q_{\text b}$ and $Q_{\text f}$) can be computed:
 \begin{equation}
    \label{eq:Qb}
    Q_{\text b} = \frac{1}{x^2}\left| \sum_{n=1}^{\infty}
    \left( 2n+1 \right)\left( -1 \right)^n\left( a_n-b_n \right)\right|^2
\end{equation}
\begin{equation}
    \label{eq:Qf}
    Q_{\text f} = \frac{1}{x^2} \left| \sum_{n=1}^{\infty}
    \left( 2n+1 \right)\left( a_n+b_n \right) \right|^2
\end{equation}
As the imaginary part of the permittivity of a sphere changes sign, its forward-to-backward scattering cross section $Q_\text f/Q_\text b$ changes drastically. This is illustrated in Figure~\ref{fig:g} for spheres with varying size (the real part of the relative permittivity is kept constant $\E'=3$). There it can be seen that while for passive spheres $\E''>0$ this ratio is of the order of unity and decreases when the size parameter becomes larger, for negative $\E''$ values, it has a very dynamic behavior and can reach very large values.
 
Another way of illustrating the backscattering enhancement is the asymmetry parameter $g$ which weighs the scattering over all the spatial directions \cite[p.~72]{Bohren-Huffman}:
\begin{align}
    g &= \left<\cos\theta\right> = \frac{Q_\text{sca}\!\left<\cos\theta\right>}{Q_\text{sca}}
    \label{eq:asy}
\end{align}
where
\begin{equation}
    Q_\text{sca}\!\left<\cos\theta\right> = 
    \frac{4}{x^2} \sum_{n=1}^\infty \frac{n(n+2)}{n+1}
    \operatorname{Re}\left\{a_n a_{n+1}^* + b_n b_{n+1}^*\right\}
    + \frac{4}{x^2}\sum_{n=1}^\infty \frac{2n+1}{n(n+1)}
    \operatorname{Re}\left\{ a_n b_n^*\right\}
\end{equation}
Positive values of $g$ indicate that the object scatters mostly into the forward hemisphere while negative values mean that the backward scattering is dominant. This is also shown in Figure~\ref{fig:g}.

While the enhanced backscattering is responsible for much of the large $Q_\text b/Q_\text f$ values (and negative $g$ values) in Figure~\ref{fig:g}, this ratio can also be locally large due to very low forward scattering cross section. For example, for a sphere with $x=1.65$ and $\E=3+\jj 2$, the back-to-forward ratio is around $17.2$ and $g\approx -0.423$. This can be connected to the {\em second Kerker condition} for which the electric and magnetic dipole Mie coefficients $a_1$ and $b_1$ are opposite complex numbers, leading to a vanishing dipole contribution in Equation~\eqref{eq:Qf} for the forward scattering efficiency. Indeed, for these values, the coefficients are
\begin{equation}
a_1 \approx 0.739 + \jj 1.16 \quad \text{and} \quad b_1 \approx -0.789 - \jj 1.10
\end{equation}

\section{Zero-extinction objects}

\subsection{Classification of passive and active scatterers}

Depending on the sign of the imaginary part of the permittivity of an object, it can be dissipative $\E''>0$, lossless (and gainless) $\E''=0$, or active $\E''<0$, keeping in mind the time-harmonic notation $\exp(\jj \omega t)$ and the electrical engineering convention $\E=\E'- \jj \E''$. Passive media are characterized by $\E''\geq 0$, including both lossless and dissipative cases. The scattering efficiency \eqref{eq:Qsca} is always positive regardless of the sign of $\E''$, but the absorption efficiency \eqref{eq:Qabs}, which has the same sign as $\E''$, can be positive, zero, or negative. This means that for active scatterers, the sign of the extinction efficiency \eqref{eq:Qext} can also vary. We can hence define three classes of active objects: positive extinction, zero extinction, and negative extinction objects, leading to the classification in Table~\ref{tab:def2}.

\begin{table}
\centering
\begin{tabular}{|l|c|c | c c c|}
\hline
& $\E''$  & & $Q_\mathrm{sca}$ & $Q_\mathrm{abs}$ & $Q_\mathrm{ext}$  \\\hline
DPE & $>0$ & dissipative (positive extinction) & $>0$ & $>0$ & $>0$ \\ \hline
LPE & $=0$ & lossless (positive extinction) & $>0$ & $=0$ & $>0$ \\ \hline
APE & $<0$ & active (positive extinction) & $>0$ & $<0$ & $>0$ \\ \hline
AZE & $<0$ & active (zero extinction) & $>0$ & $=-Q_\mathrm{sca}$ & $=0$ \\ \hline
ANE & $<0$ & active (negative extinction) & $>0$ & $<0$ & $<0$ \\ \hline
 
\end{tabular}
\caption{\label{tab:def2} 
Five classes of scatterers (two passive and three active), determined by the signs of the absorption and extinction efficiencies.}
\end{table}

\subsection{Zero-extinction scatterers}

\begin{figure}[htbp]
    \centering
    \subfloat[$\E=1.1$]{\includegraphics[width=45mm]{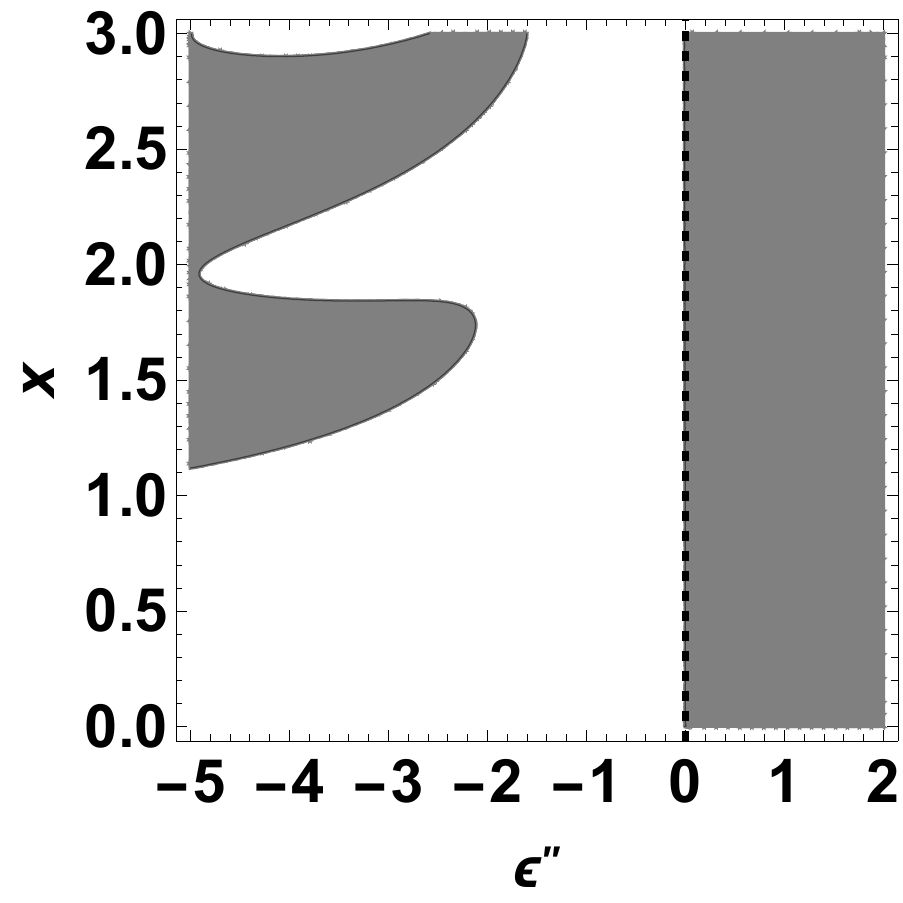}\hspace{1mm}}
    \subfloat[$\E=4$]{\includegraphics[width=45mm]{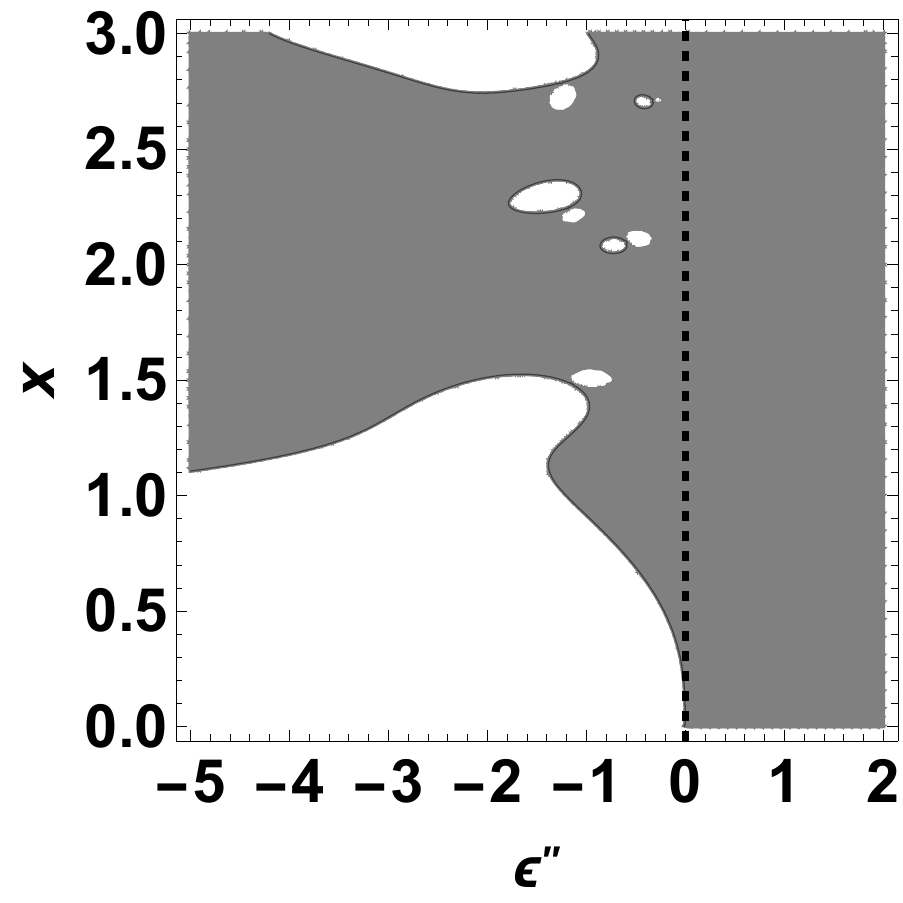}\hspace{1mm}}
    \subfloat[$\E=7$]{\includegraphics[width=45mm]{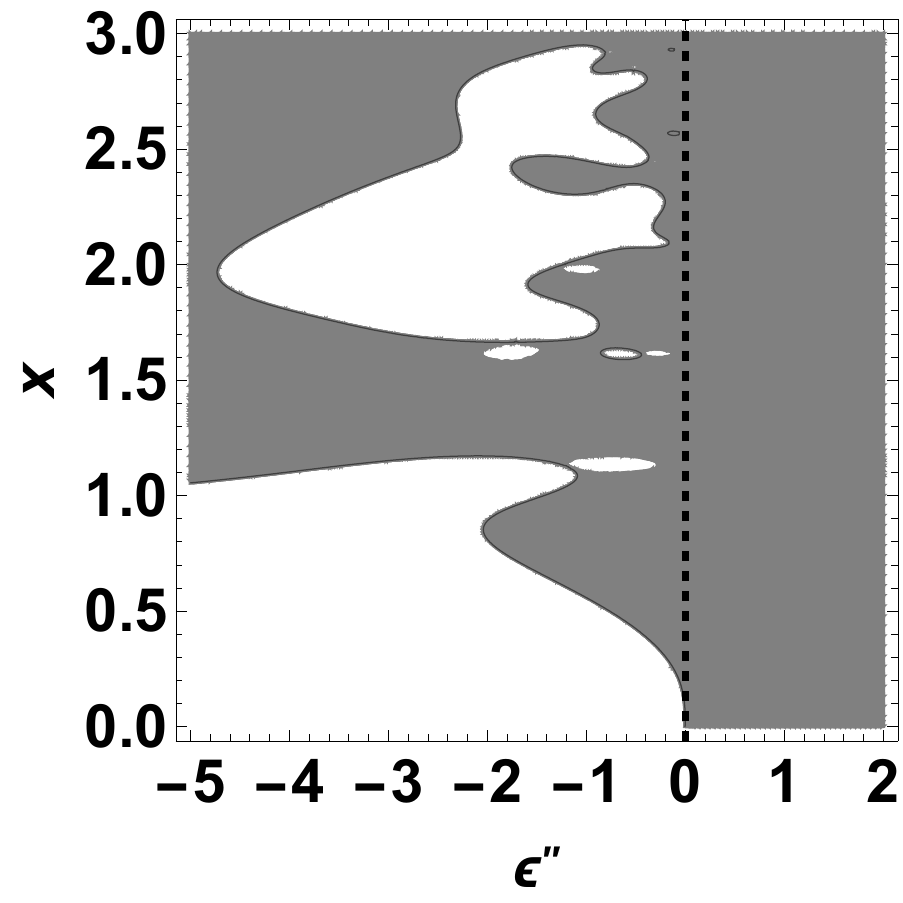}\hspace{1mm}}
    \caption{\label{fig:zeo} The sign of the extinction efficiency of dielectric spherical scatterers depicted as gray (APE and DPE, $Q_\text{ext}>0$) and white (ANE, $Q_\text{ext}<0$) in the plane with axes of the imaginary part of the relative permittivity $\varepsilon''$  and the size parameter of the sphere $x$. The dashed vertical black lines at $\E''=0$ give the lossless/gainless case LPE. Note the rich structure of extinction variation on the active side ($\varepsilon''<0$).}
\end{figure}

The three efficiencies, as Table~\ref{tab:def2} shows, can combine with different signs when the parameters of the object change. While extinction is always positive for passive scatterers, the active side is very dynamic as the sign of extinction varies. Figure~\ref{fig:zeo} displays this behavior within the parametric space of the scatterers (size and the complex permittivity), and illustrates how, on the active side, the three scatterer types (APE, AZE, and ANE) populate themselves into a fractal-like varying landscape. In other words, zero-extinction (AZE) objects can be found in several locations (on the boundaries of the white islands of Figure~\ref{fig:zeo}) in this three-parameter $(x,\E',\E'')$ space.


\subsection{Optical theorem} 

The so-called \emph{optical theorem} in the scattering theory has a long history which starts already from 150 years ago, with the studies by John William Strutt (who later rose in peerage as Lord Rayleigh) \cite{Rayleigh,Newton-opticaltheorem}. This fundamental theorem connects the extinction efficiency of an object to its forward scattering characteristics.

Using the notation in Bohren and Huffman \cite[p.~112]{Bohren-Huffman}, the extinction cross section of a sphere is connected to the scattered field into the forward direction ($S(0^\circ)$) as 
\begin{equation}
    C_\text{ext} = \frac{4\pi}{k^2}\text{Re}\{S(0^\circ\}
\end{equation}
where
\begin{equation}
S(0^\circ) = \frac{1}{2} \sum_n (2n+1) (a_n+b_n)
 \end{equation}
This gives us a form of the optical theorem:
\begin{equation}\label{eq:S0}
    Q_\text{ext} = \frac{4}{x^2}\text{Re}\{S(0^\circ)\}
\end{equation}

For passive scatterers, the connection has been established in the past literature. Taking arbitrary parameters, like, for example $\E=2.3-\jj 1.2$ and $x=3.4$, we have
\begin{equation}
Q_\text{ext} =2.76096\quad \text{and} \quad \frac{4}{x^2}S(0^\circ)= 2.76096 +\jj \,0.14832
\end{equation}

The optical theorem seems particularly relevant in the special case of AZE scatterers for which the extinction cross section vanishes. However, this does not mean that the forward scattering should vanish; only the real part of $S(0^\circ)$ according to \eqref{eq:S0}. Let us confirm this result numerically by taking an AZE sphere $\E=2+\jj$ with $x=2.69156$. In this case we find that 
\begin{equation}
    Q_\text{ext} \approx 0 \quad \text{and} \quad\frac{4}{x^2}S(0^\circ) \approx \jj\, 3.40627
\end{equation}
in agreement with the optical theorem.

\section{\label{sec:numerics} Morphological effects on scattering response by active particles}

\begin{figure}[tbp]
\centering
  \subfloat[$Q_\text{sca}$]{\includegraphics[width=0.43\textwidth]{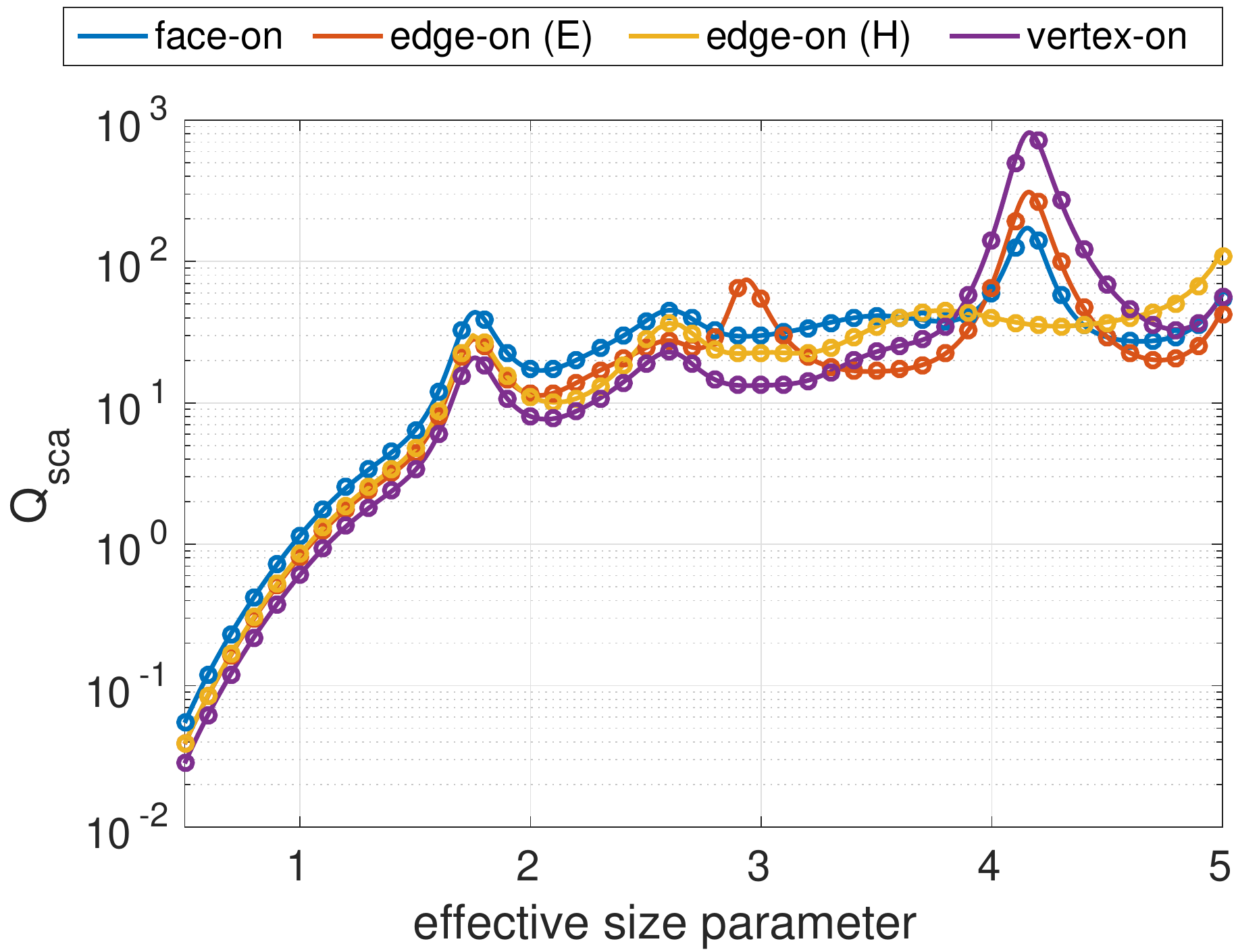}}\quad
  \subfloat[$Q_\text{abs}$]{\includegraphics[width=0.43\textwidth]{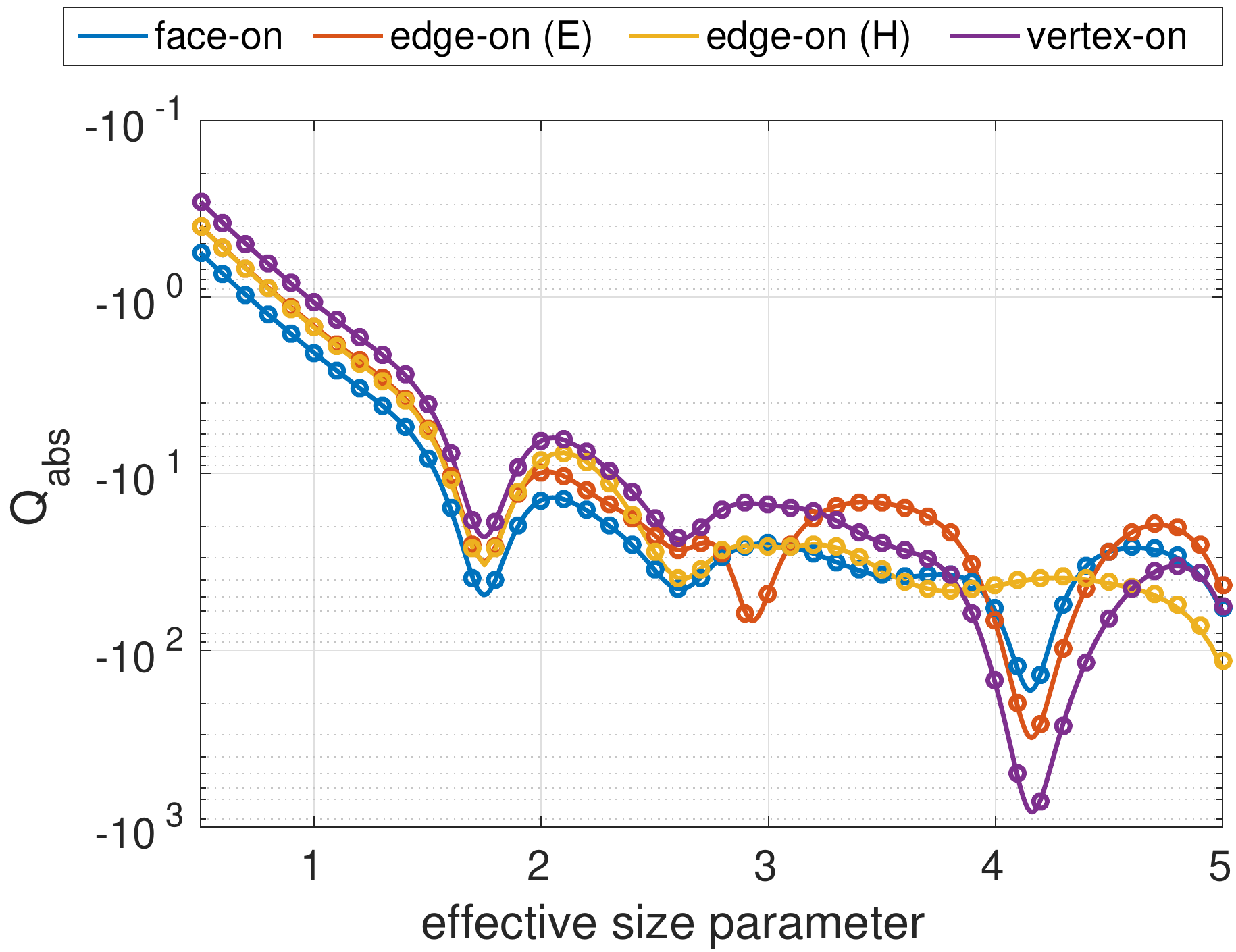}} 
\caption{Scattering (a) and absorption (b) efficiency with COMSOL (solid lines) and MoM (circles) in logarithmic scale for a cube with $E = 3+1.5\jj$ and with four different incident waves. The cube has the same volume as a sphere with size parameter $x=ka$.}
\label{fig:cube_Qsca_Qabs}
\end{figure}
\begin{figure}[tbp]
\centering
\includegraphics[width=0.8\textwidth]{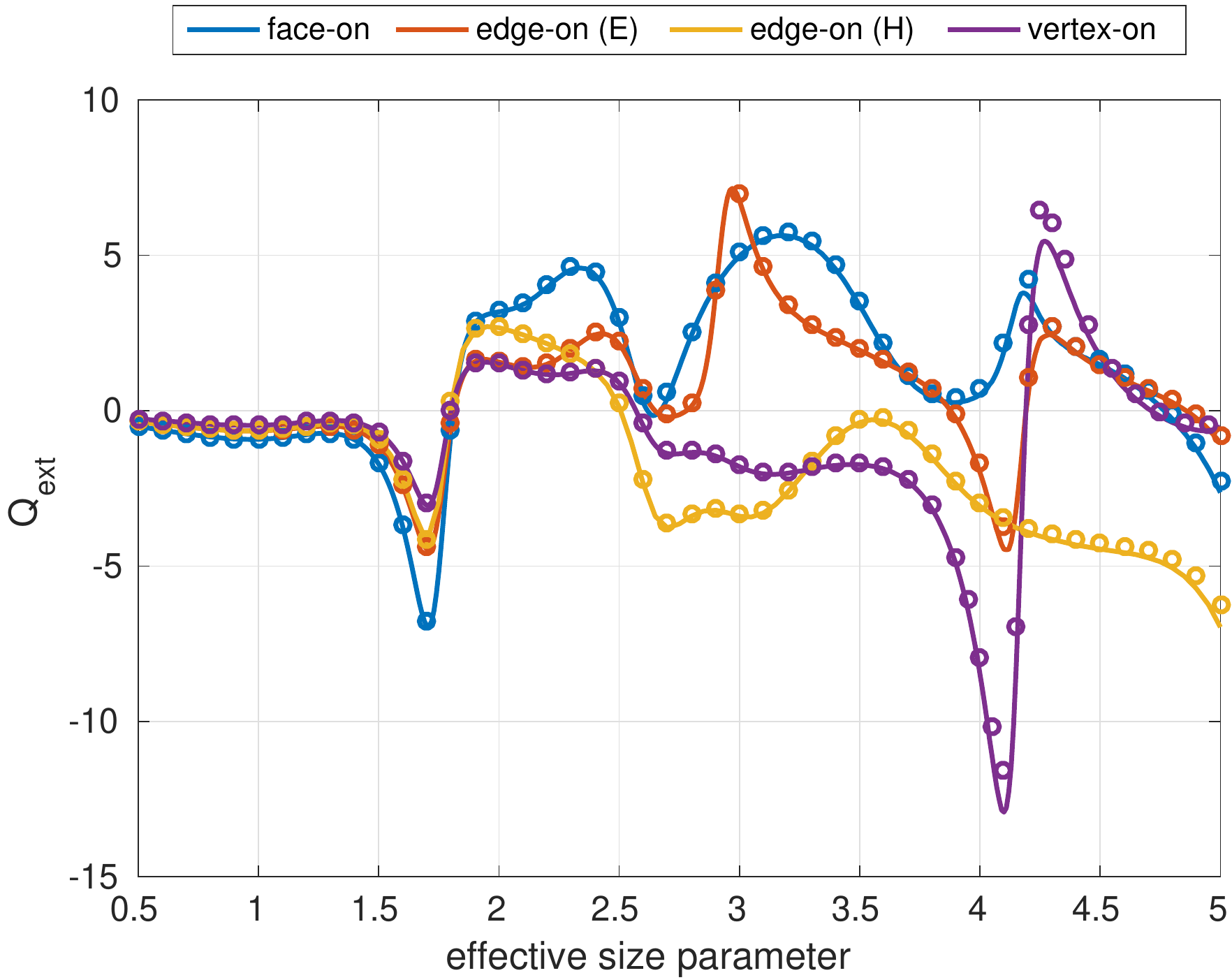} 
\caption{Extinction efficiency with COMSOL (solid lines) and MoM (circles) for a cube with $\E = 3+1.5\jj$ and with four different incident waves, as in Figure~\ref{fig:cube_Qsca_Qabs}.}
\label{fig:cube_Qext}
\end{figure}

So far the scattering response of spherical objects has been investigated. In this section we study qualitative effects of the shape of an active particle on the response of scattering and extinction characteristics. We proceed with another symmetric object, a cube, in which case we consider four different incident fields: 
\begin{itemize}
\item {\it Face-on}: The propagation direction is perpendicular to one of the faces of the cube and the incident electric field is parallel to one of the edges of the cube.
\item {\it Edge-on (E)}: The propagation direction is towards one of the edges of the cube with $45^\circ$ angle to one of the faces of the cube, and the incident electric field is parallel to that edge.
\item {\it Edge-on (H)}: The same as Edge-on (E), but the incident magnetic field is parallel to the edge.
\item {\it Vertex-on}: The propagation direction is towards one of the vertices of the cube along the cube diagonal.
\end{itemize}
Figures \ref{fig:cube_Qsca_Qabs} and \ref{fig:cube_Qext} show the scattering, absorption and extinction efficiency for a cube having the same volume as a sphere with size parameter $x=ka$. The efficiency is normalized using the geometrical cross section of the cube, and given for the four incident fields mentioned above. The results computed with both COMSOL and MoM are shown. The scattering response (efficiency) and the geometrical cross section depend on the incident field propagation direction and polarization, except in the vertex-on case where the solutions are independent on the incident field polarization. 

\begin{figure}[tbp]
    \newlength\rcwidth
    \setlength\rcwidth{30mm}
    \centering
    \parbox{\rcwidth}{\includegraphics[width=\rcwidth]{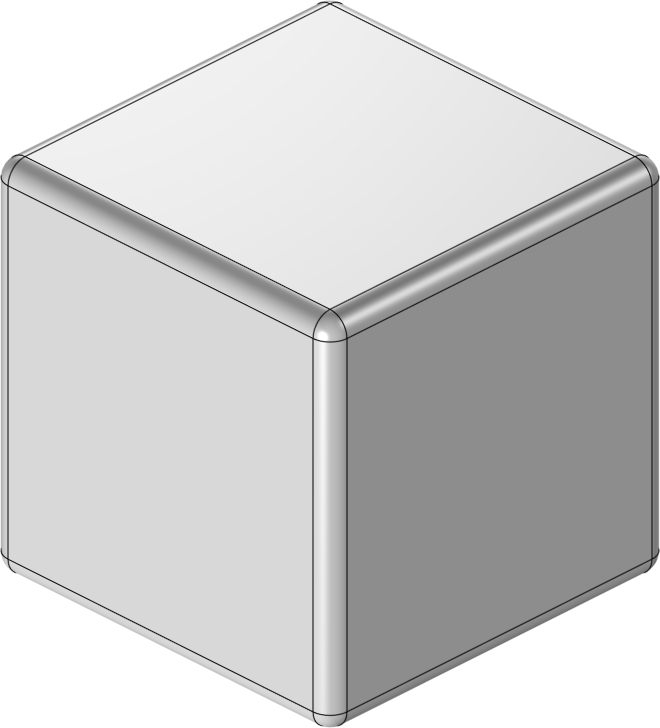}\\
        \centerline{$r_0=L/20$}}\quad
    \parbox{\rcwidth}{\includegraphics[width=\rcwidth]{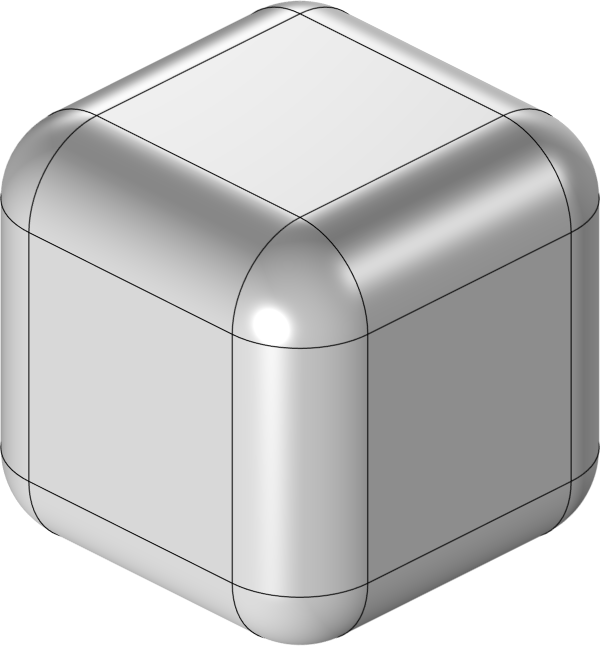}\\
        \centerline{$r_0=L/5$}}\quad
    \parbox{\rcwidth}{\includegraphics[width=\rcwidth]{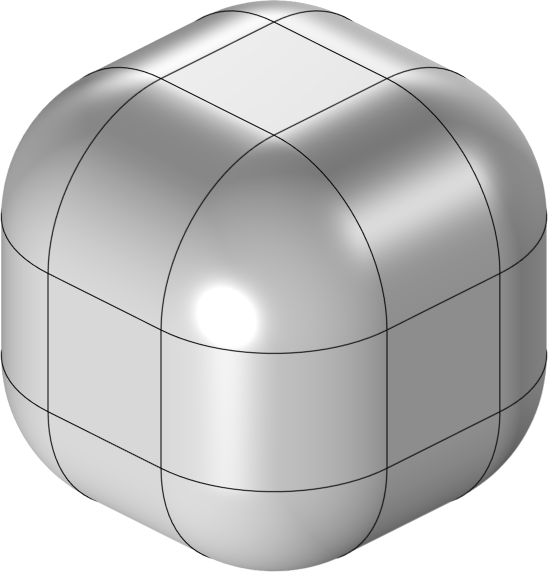}\\
        \centerline{$r_0=L/3$}}\quad
    \parbox{\rcwidth}{\includegraphics[width=\rcwidth]{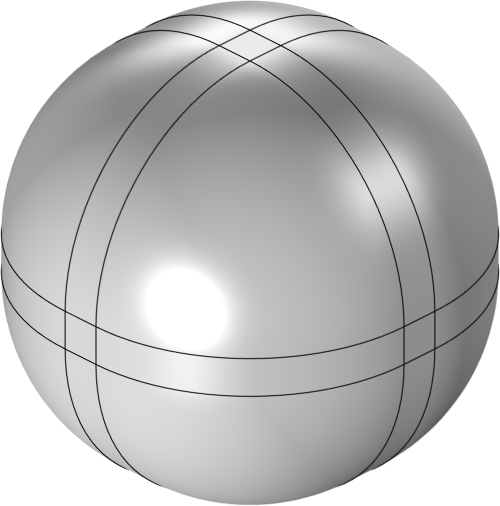}\\
        \centerline{$r_0=L/2.2$}}
    \caption{Geometry of a rounded cube with side length $L$ and different fillet radius $r_0$. From left to right a smooth transition from a slightly rounded cube to almost a sphere.}
    \label{fig:rcube_geom}
\end{figure}

\begin{figure}[htp]
  \centering
  \subfloat[$Q_\text{sca}$]{\includegraphics[width=0.43\textwidth]{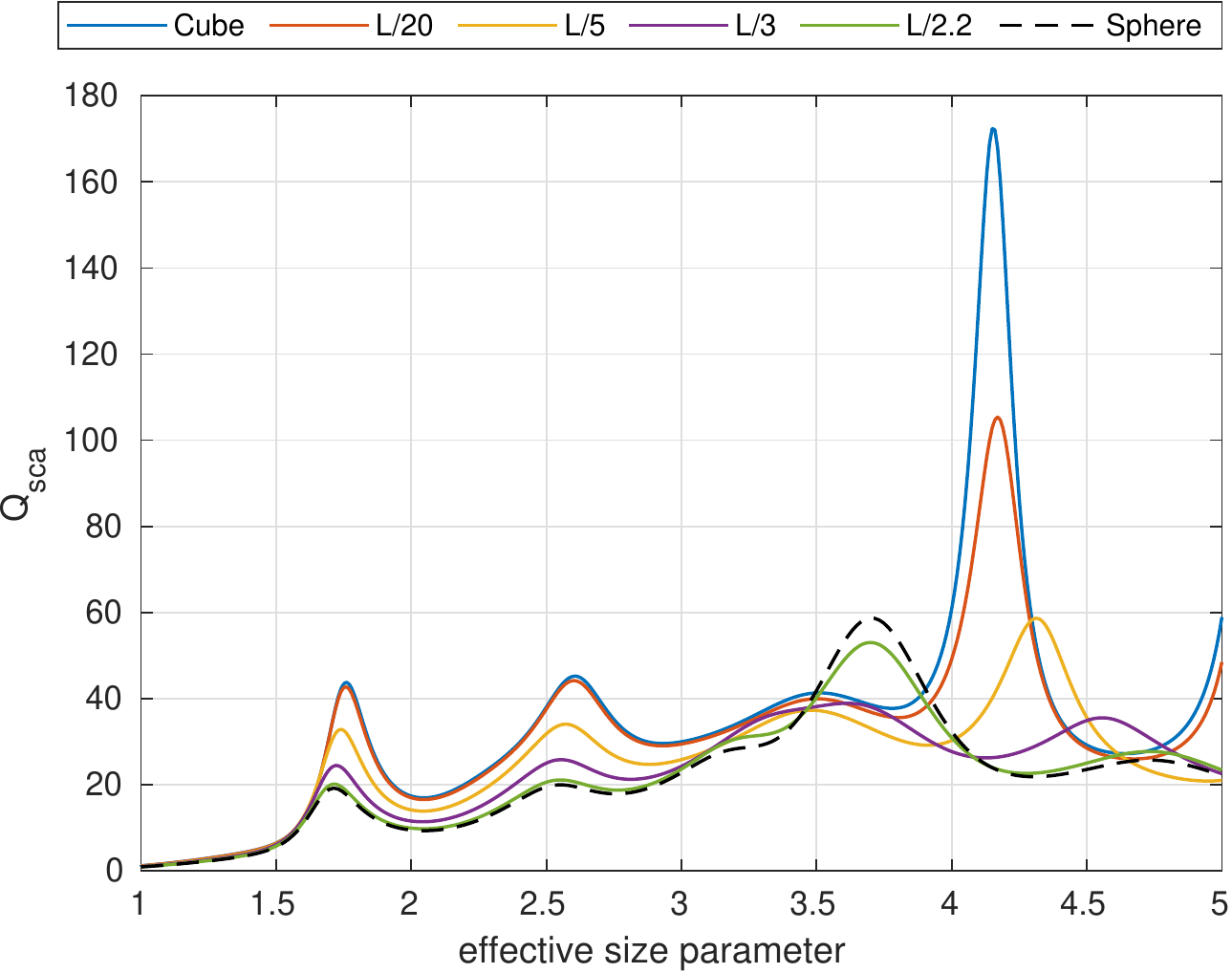}}\quad
  \subfloat[$Q_\text{abs}$]{\includegraphics[width=0.43\textwidth]{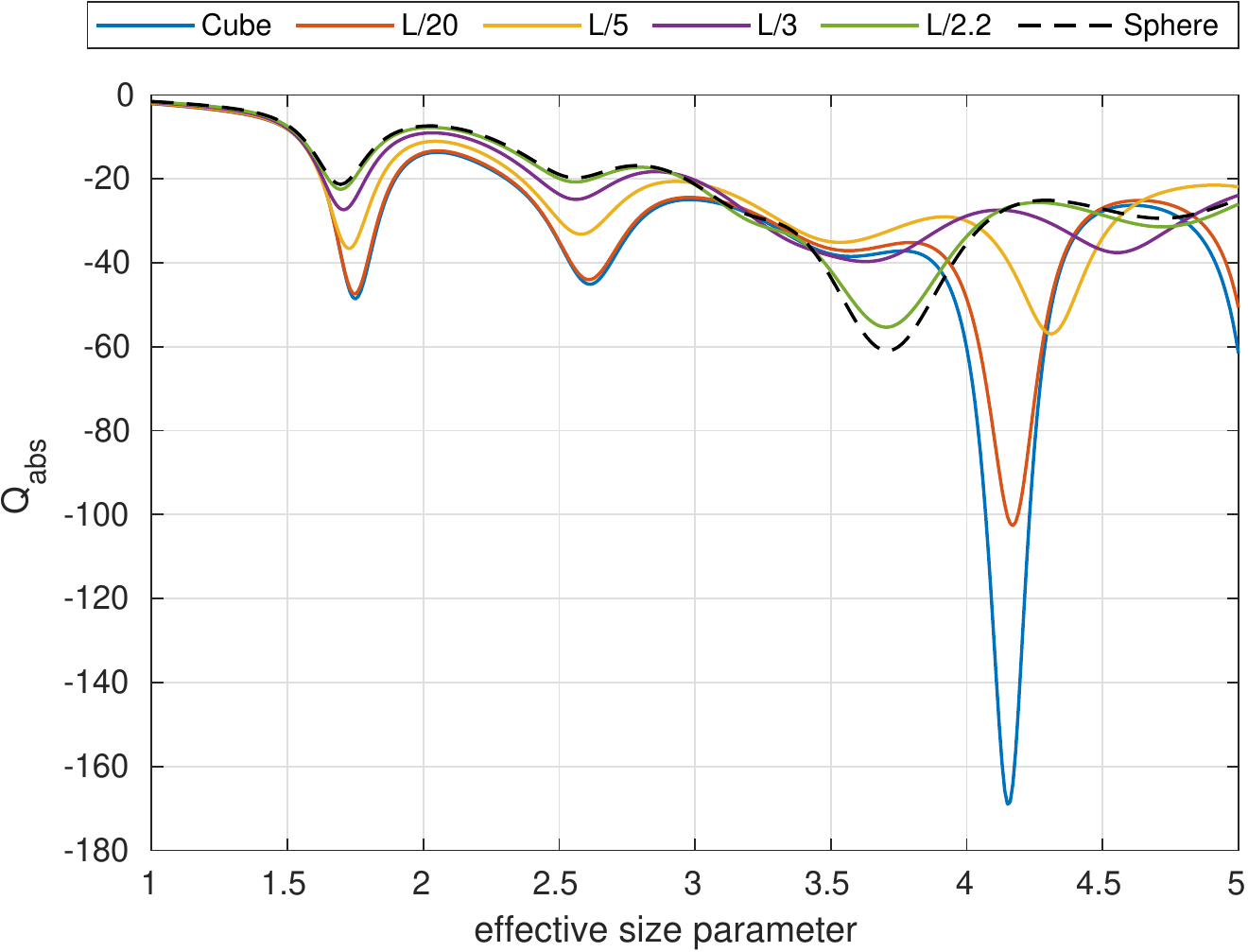}}
  \caption{Scattering (a) and absorption (b) efficiency for an active cube with $\E=3+\jj 1.5$ transforming into a sphere and face-on polarization. The geometry of the intermediate steps are shown in Figure~\ref{fig:rcube_geom}. For a sphere the results are computed with Mie series, in other cases COMSOL solutions are shown. All scatterers have the same volume and the efficiency is normalized using the geometrical cross section of each object.}
    \label{fig:rcube_Qsca_Qext}
\end{figure}

To study the effect of the shape of an active particle on the response of scattering and extinction characteristics, the transformation of a sharp cube into a sphere is considered next. The geometries  (rounded cubes) used in this study are illustrated in Figure~\ref{fig:rcube_geom}. Figure \ref{fig:rcube_Qsca_Qext} shows the scattering and absorption efficiency with face-on polarization. At the first two resonances (maxima of the efficiency) close to $x=1.7$ and $x=2.6$, the effect of this transformation on the efficiency is relative smooth, while for other resonances at larger $x$ values the situation is more complicated.

\begin{figure}[htp]
\centering
\subfloat[$x=1.71$]{\includegraphics[trim=0 25 25 25, clip, width=0.5\textwidth]{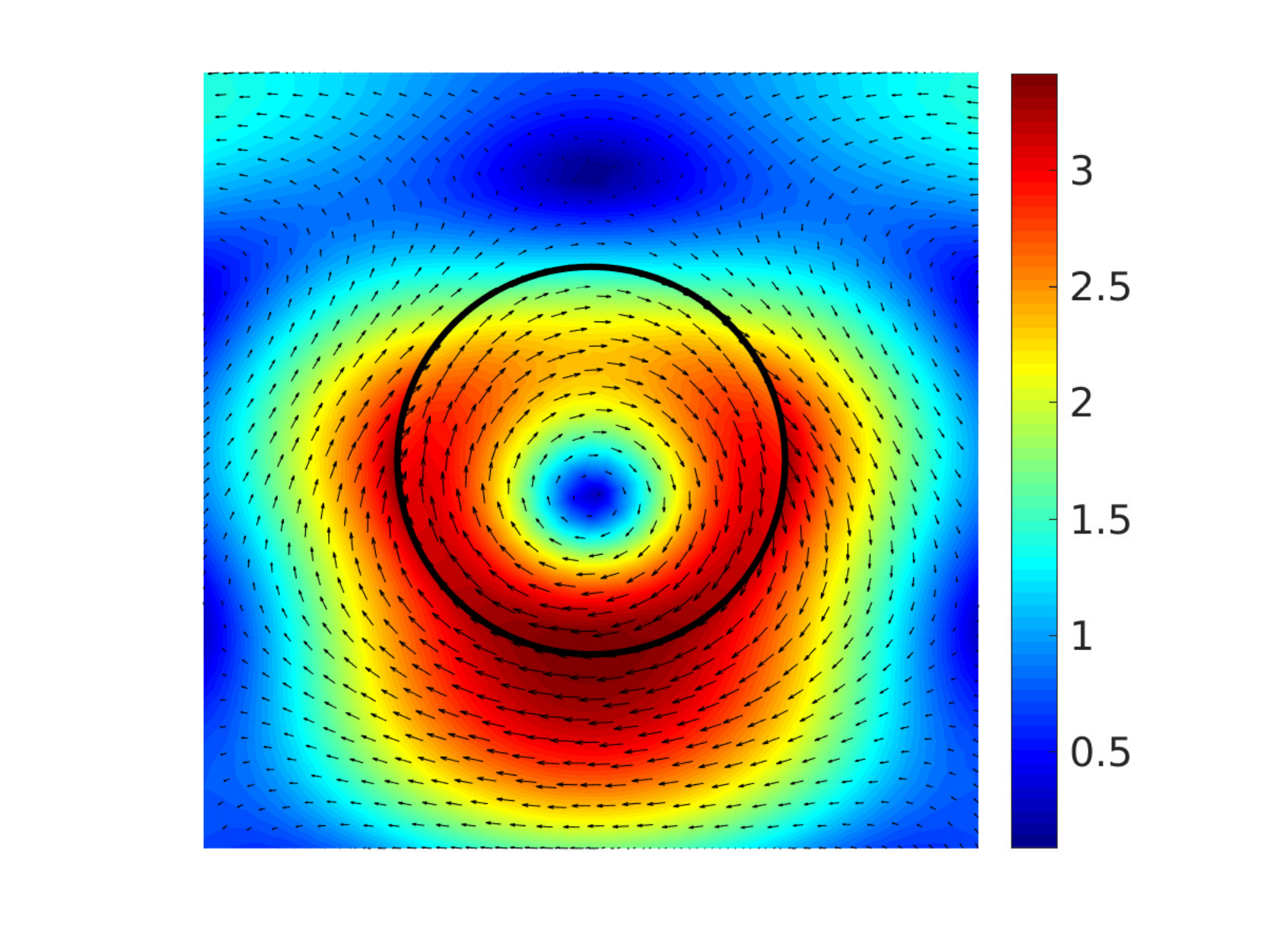}} 
\hspace{-5mm}
\subfloat[$x=2.55$]{\includegraphics[trim=25 25 0 25, clip, width=0.5\textwidth]{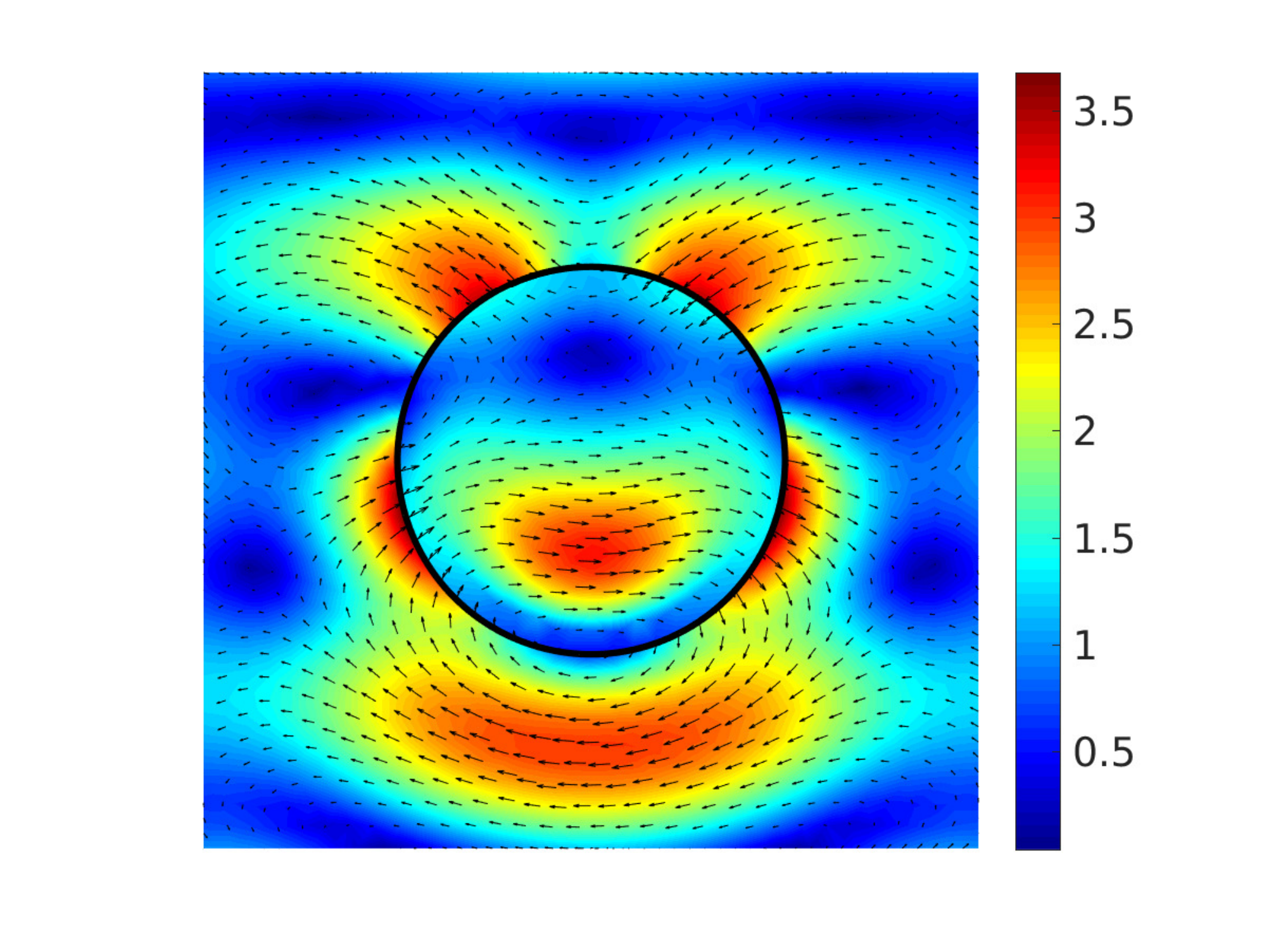}}
\hspace{-5mm}
\subfloat[$x=3.70$]{\includegraphics[trim=0 25 25 25, clip, width=0.5\textwidth]{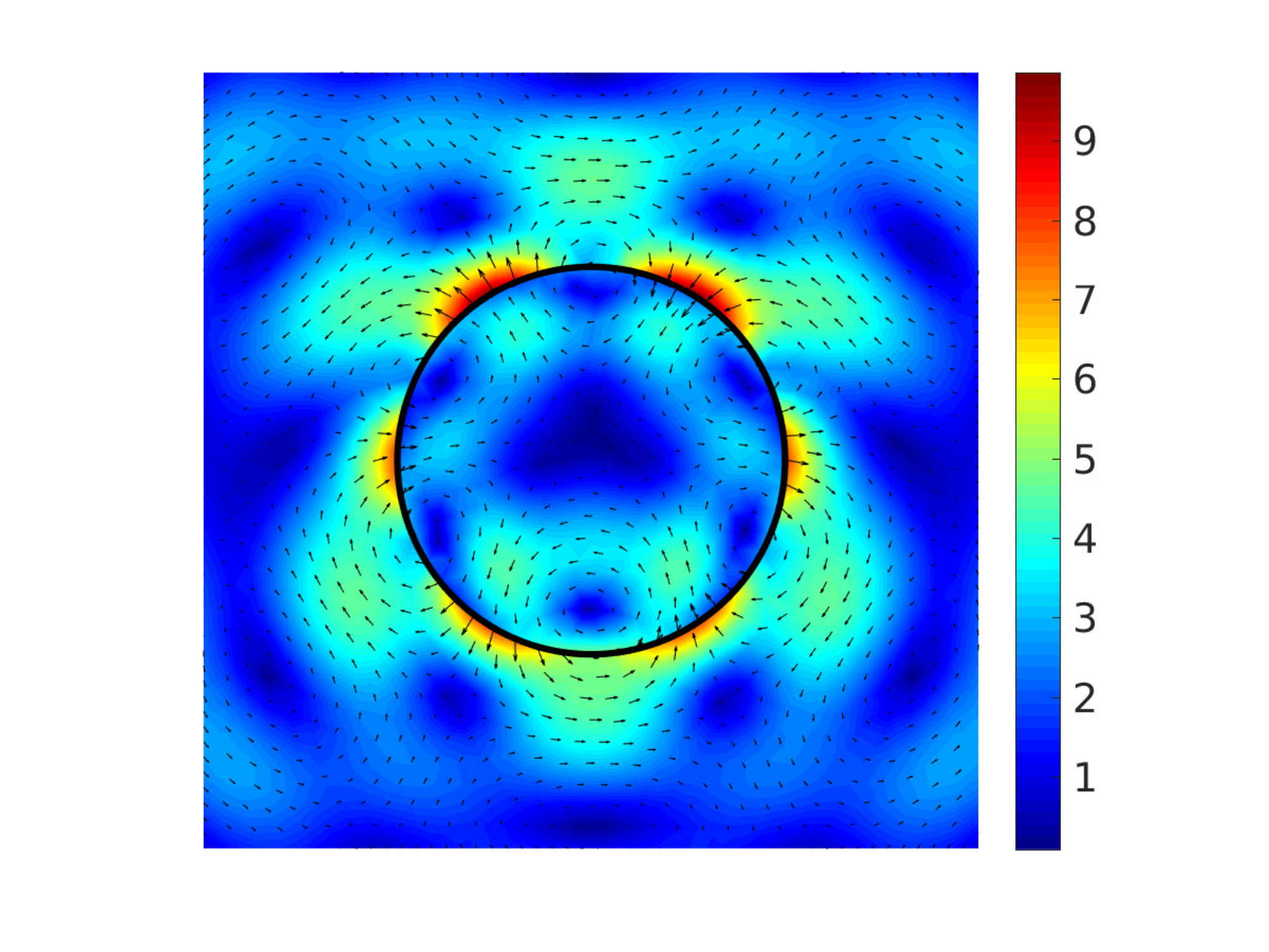}}
\hspace{-5mm}
\subfloat[$x=4.75$]{\includegraphics[trim=25 25 0 25, clip, width=0.5\textwidth]{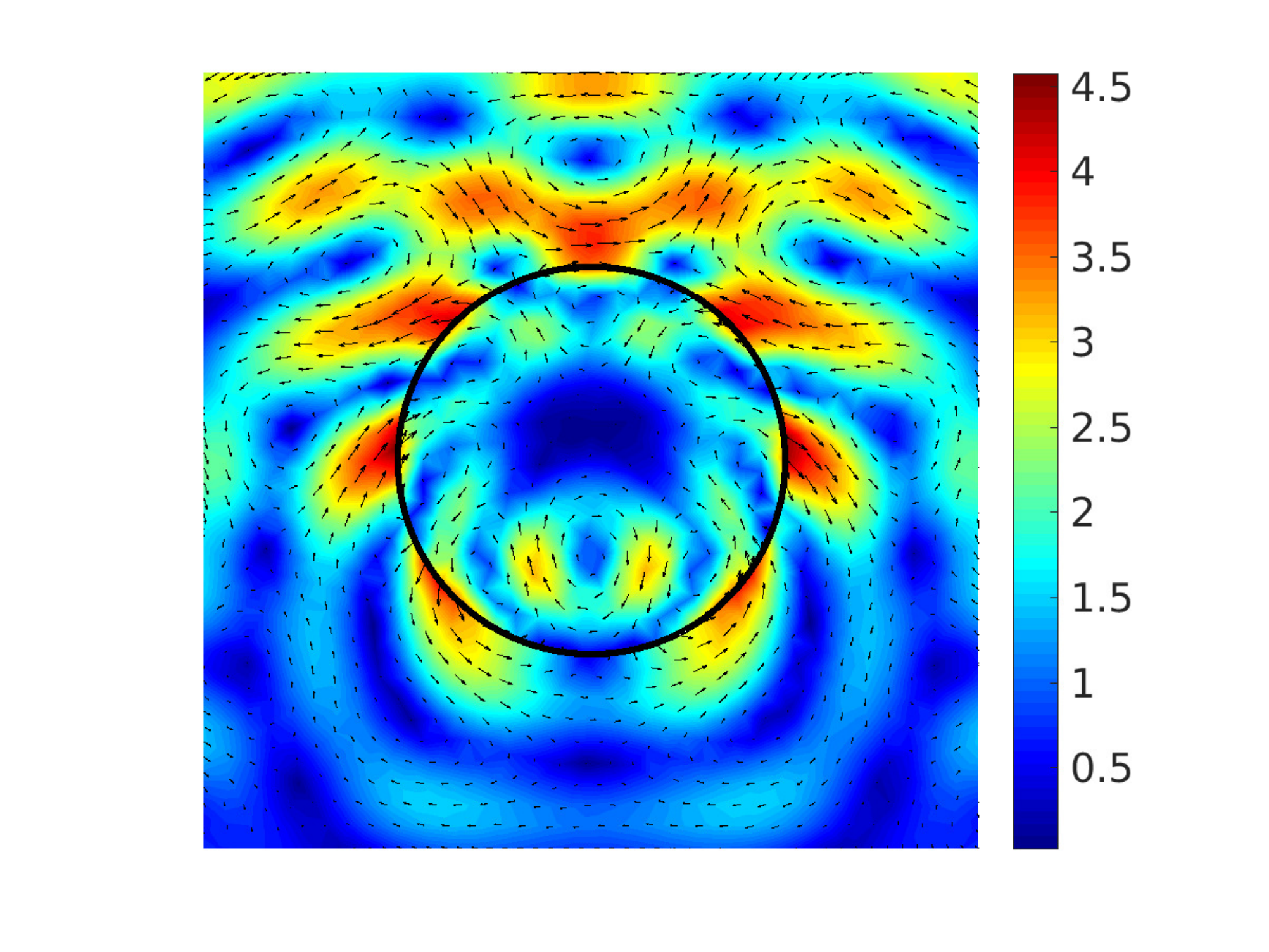}}
\caption{\label{fig:Efield} Electric field (computed with Mie series) at the $xz$-plane for an active sphere with $\E = 3+1.5\jj$ and with different size parameters. The incident field is propagating along the $+z$ axis and incident electric field is $x$ polarized. Color indicates magnitude (color scale varies in the figures) and arrows show the direction of the field. In the figures, $x$-axis is horizontal and $z$-axis vertical.}
\end{figure}

\begin{figure}[htp]
\centering
\subfloat[$x=1.75$]{\includegraphics[trim=0 25 25 25, clip, width=0.5\textwidth]{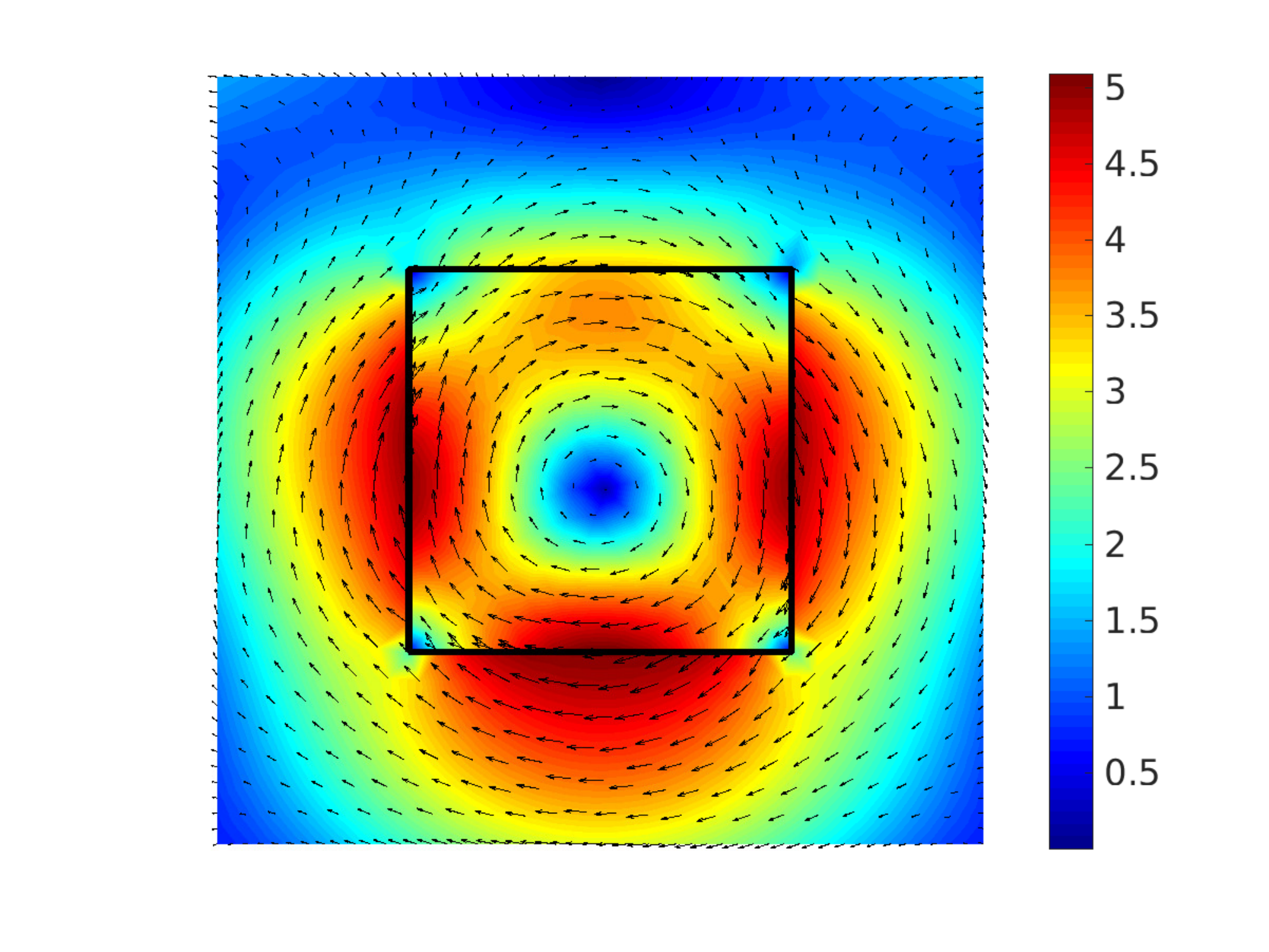}} 
\hspace{-5mm}
\subfloat[$x=2.60$]{\includegraphics[trim=0 25 25 25, clip, width=0.5\textwidth]{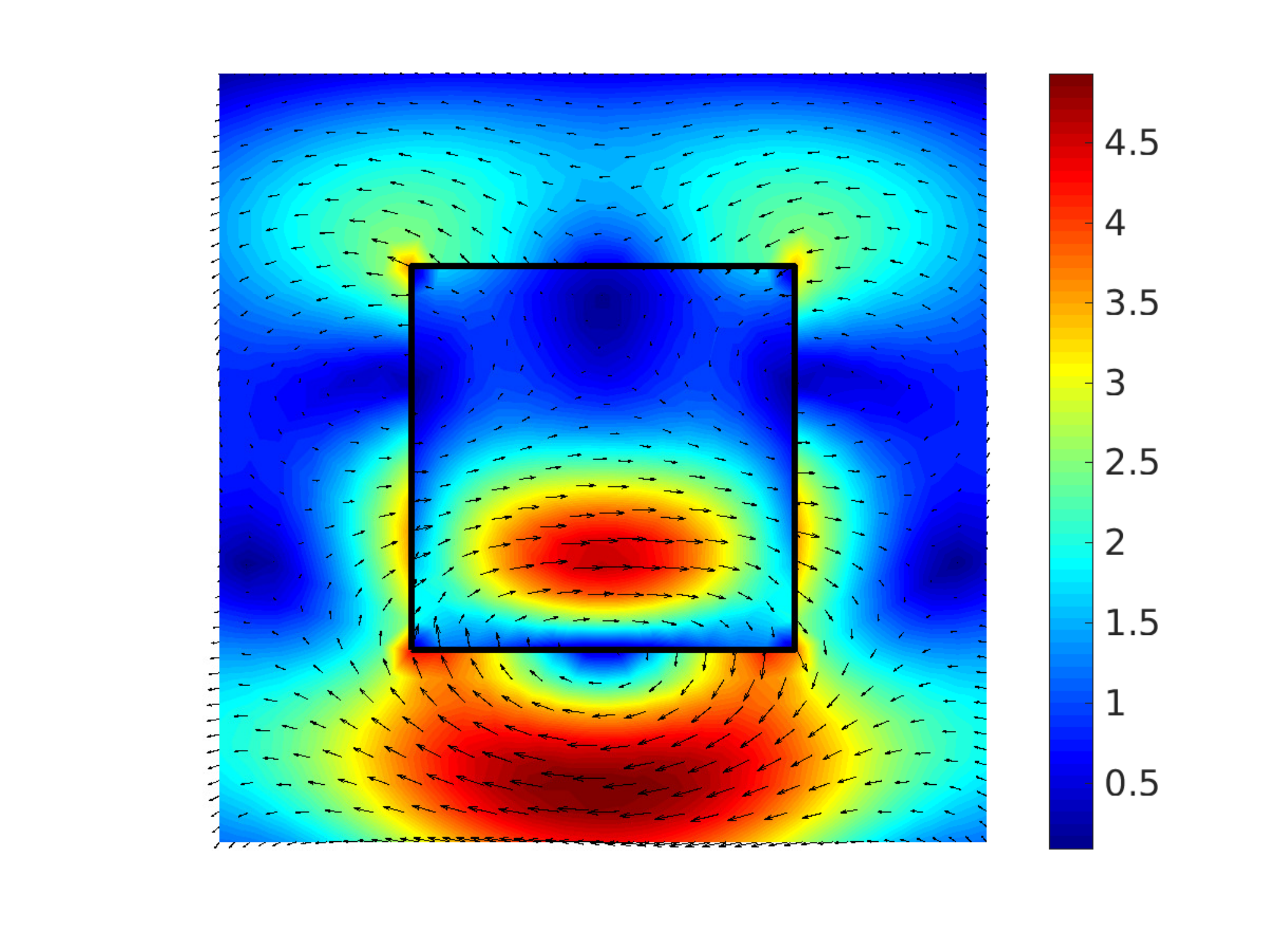}}
\hspace{-5mm}
\subfloat[$x=3.50$]{\includegraphics[trim=0 25 25 25, clip, width=0.5\textwidth]{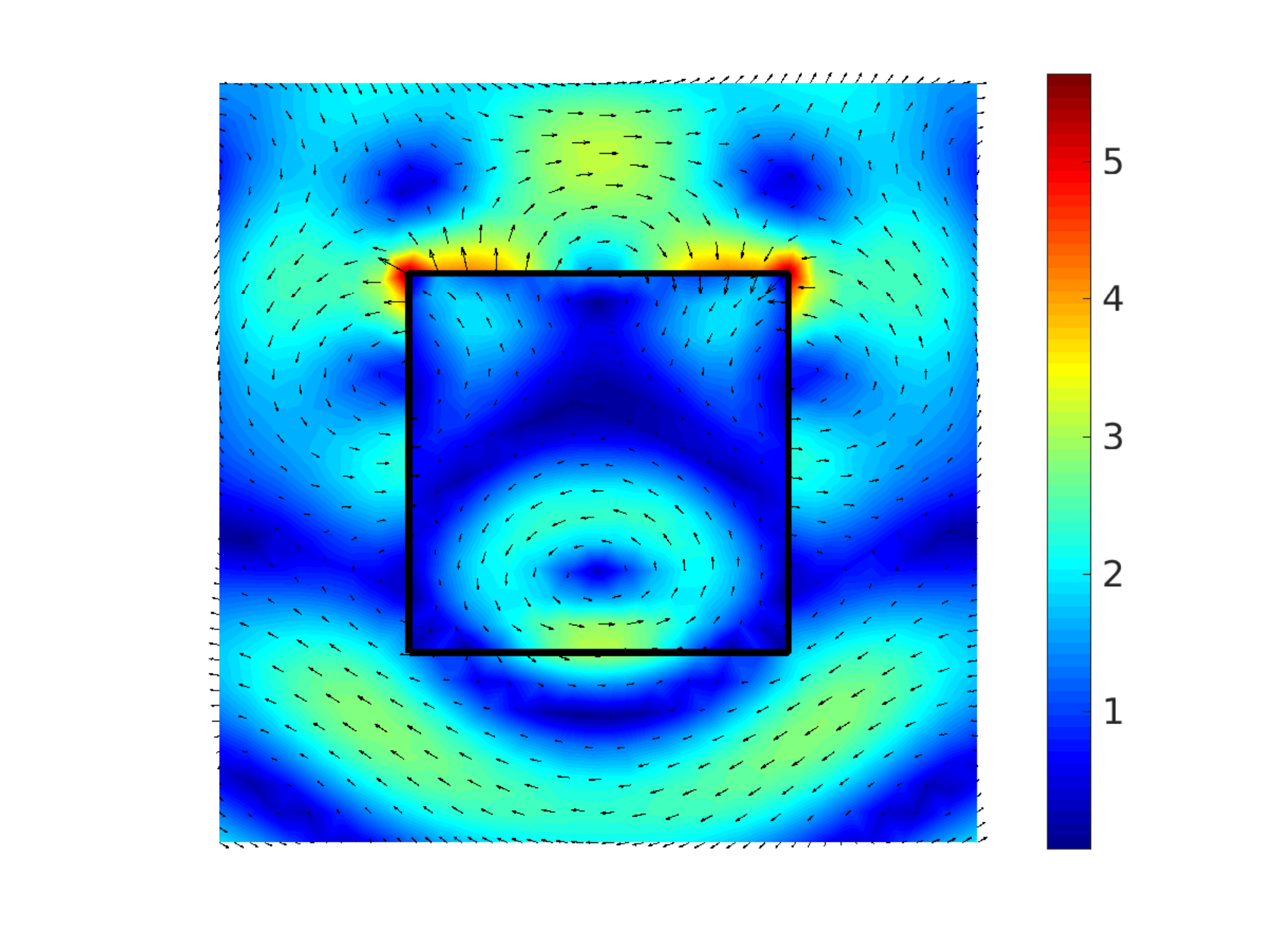}}
\hspace{-5mm}
\subfloat[$x=4.15$]{\includegraphics[trim=0 25 25 25, clip, width=0.5\textwidth]{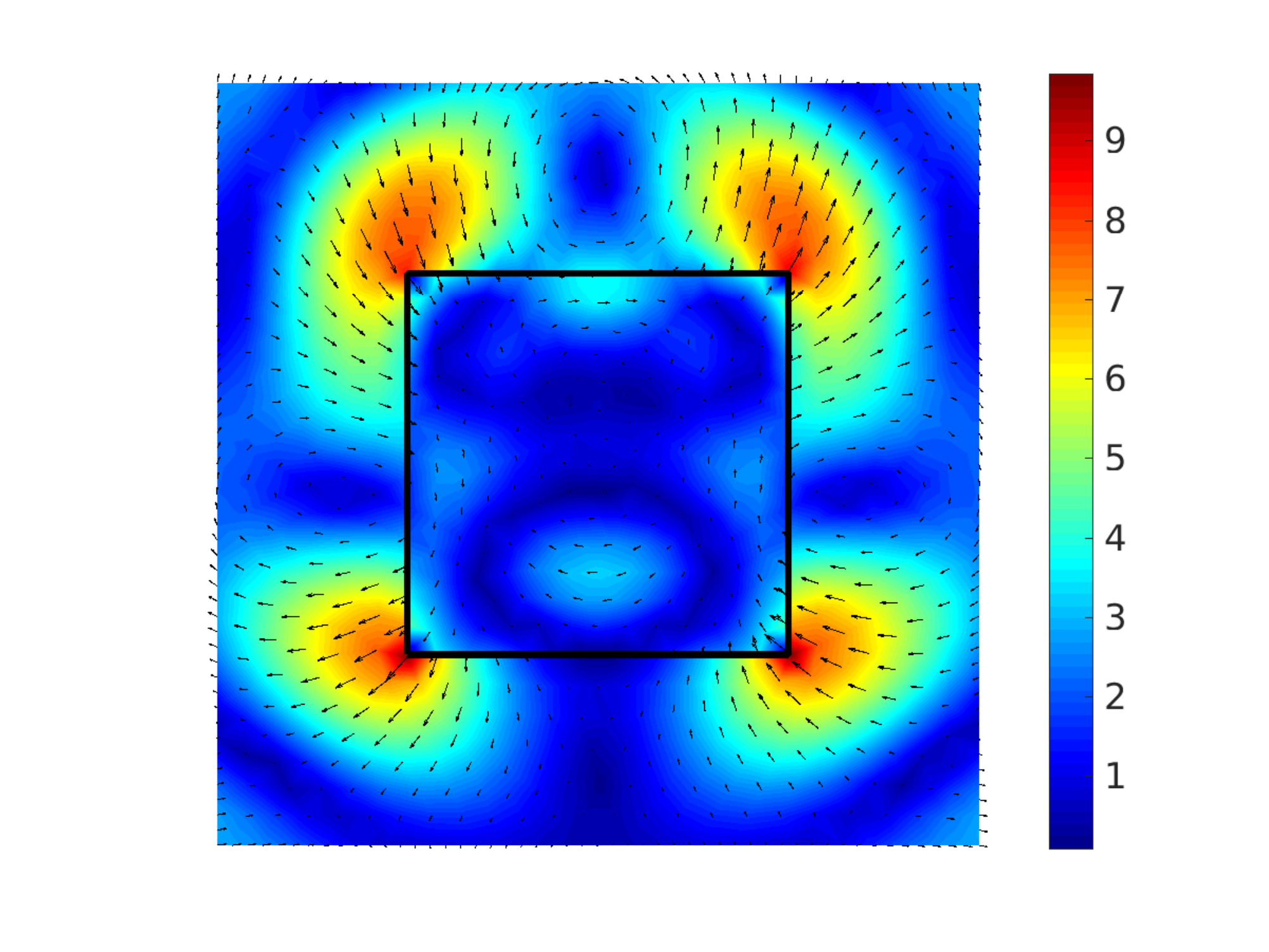}}
\caption{\label{fig:Efield_cube} Electric field (computed with MoM) at the xz-plane for an active cube with $\E = 3+1.5\jj$ and with different size parameters. The incident field is propagating along the $+z$ axis and incident electric field is $x$ polarized. In the figures, $x$-axis is horizontal and $z$-axis vertical.}
\end{figure}

To illustrate the field behavior near the resonances, we take a look at the near-field solutions of active particles.
Figures \ref{fig:Efield} and \ref{fig:Efield_cube} show the electric field at the $xz$-plane for an active sphere and cube with $\E = 3+1.5\jj$. In both cases the field is plotted for four different size parameters corresponding to the maxima of the scattering efficiency in Figure \ref{fig:comparison} and in Figure \ref{fig:cube_Qsca_Qabs}, respectively. The incident field is a $+z$-propagating plane wave with an $x$-polarized electric field.

Clearly, the first two resonances of the sphere and cube are due to similar field solutions. This may also be concluded from Figure \ref{fig:rcube_Qsca_Qext} where the first two maxima of $Q_{\rm sca}$ for a sphere and cube are relative close to each other, at $x=1.71$ and $x=2.55$ (sphere) and at $x=1.75$ and $x=2.60$ (cube). Also at the third resonances ($x=3.7$ for a sphere and $x=3.5$ for a cube) we may recognize some similarities in the field solutions. The situation with the fourth resonances, however, is more complicated and a clear correspondence between the field solutions is not recognized. We note that for a cube with $x=4.15$ (Figure \ref{fig:Efield_cube}(d)) the fields are strongly localized to the vertices and exactly the same field solution may not exist for a sphere. 



\section{Conclusion}

The scattering response of material particles to electromagnetic and optical wave excitation depends strongly on their geometry and medium constitution. This paper focused on the absorption, scattering, and extinction of non-magnetic objects whose permittivity is isotropic but active (gainy), in other words, the imaginary part of the permittivity has the opposite sign than particles that are dissipative. Since the electromagnetic response of the scatterer needs to be described by absorption and extinction, in addition to the scattering, we arrived at an interesting classification of the objects (DPE, LPE, APE, AZE, ANE), depending on the sign (positive, zero, or negative) of the extinction cross section. 

Lorenz--Mie theory was applied in most of the computations. Among the interesting results was the characteristic property of electromagnetically active particles that they tend to direct scattering into the backward half plane and their strong back-to-front scattering ratio. Nevertheless, there are strong theoretical identities, like the extinction paradox and optical theorem, that remain valid even if the scatterer is active. These were discussed. Finally, using numerical approaches (Method of Moments and finite-element--based codes), the focus was on scatterer shapes differing from sphere (sharp and rounded cubes). The scattering spectrum of the particles and the position of the resonances evolves along with the change of the geometry from a sphere towards a cube, leading to a possibility for sensing the particle shape using the data of its electromagnetic scattering response. 

%
%
%


\end{document}